\documentclass[AMA,STIX1COL]{WileyNJD-v2}
\usepackage[utf8]{inputenc}
\usepackage[T1]{fontenc}
\usepackage{moreverb}
\usepackage{anyfontsize}

\newcommand\BibTeX{{\rmfamily B\kern-.05em \textsc{i\kern-.025em b}\kern-.08em
T\kern-.1667em\lower.7ex\hbox{E}\kern-.125emX}}

\articletype{Journal}%
    
\received{<day> <Month>, <year>}
\revised{<day> <Month>, <year>}
\accepted{<day> <Month>, <year>}

\usepackage{tablefootnote}
\usepackage{booktabs,threeparttable}
\usepackage{subcaption}
\usepackage{adjustbox}
\usepackage{amssymb}

\usepackage{amsmath}
\usepackage{comment}
\usepackage{xcolor}
\usepackage{xspace}
\usepackage{multirow}
\usepackage{tcolorbox}
\usepackage{footmisc}
\usepackage{array}
\usepackage{enumitem}

\newtcolorbox{answerbox}[1]{
    enlarge top by=10pt,
    width=\linewidth,
    halign=justify,
    colframe=black,
    colback=white,
    arc=0pt,
    outer arc=0pt,
    boxrule=0.5pt,
    title=#1
}

\usepackage{graphicx}

\usepackage{amsmath}
\usepackage{bm}

\usepackage{array}

\usepackage{url}

\newcommand\nw[1]{\textcolor{black}{#1}}

\newcommand\benchSuite{\textit{VMBS}\xspace}

\newcommand\fullindicator{Variability Indicator\xspace}
\newcommand\indicator{\textit{VI}\xspace}
\newcommand\indicators{\textit{VI}s\xspace}

\begin{document}
\title{A Multi-faceted  Analysis of the Performance Variability \\of Virtual Machines}

\author{Luciano Baresi}
\author{Tommaso Dolci}
\author{Giovanni Quattrocchi}
\author{Nicholas Rasi}

\address{Dipartimento di Elettronica, Informazione e Bioingegneria\\Politecnico di Milano, Milan, Italy}

\authormark{Baresi, Dolci, Quattrocchi, Rasi}
%\authormark{Baresi, Quattrocchi, Rasi}

\corres{*Giovanni Quattrocchi, Corresponding address. \email{giovanni.quattrocchi@polimi.it}}

\abstract[Abstract]{
Cloud computing and virtualization solutions allow one to rent the virtual machines (VMs) needed to run applications on a pay-per-use basis, but rented VMs do not offer any guarantee on their performance. Cloud platforms are known to be affected by performance \textit{variability}, but a better understanding is still required.

This paper moves in that direction and presents an in-depth, multi-faceted study on the performance variability of VMs.  Unlike previous studies, our assessment covers a wide range of factors: 16 VM types from 4 well-known cloud providers, 10 benchmarks, and 28 different metrics.
We present four new contributions. First, we introduce a new benchmark suite (\benchSuite) that let researchers and practitioners systematically collect a diverse set of performance data. Second, we present a new indicator, called \textit{Variability Indicator}, that allows for measuring variability in the performance of VMs. Third, we illustrate an analysis of the collected data across four different dimensions: \textit{resources}, \textit{isolation}, \textit{time}, and \textit{cost}. Fourth, we present multiple predictive models based on Machine Learning that aim to forecast future performance and detect time patterns.
Our experiments provide important insights on the resource variability of VMs, highlighting differences and similarities between various cloud providers.
To the best of our knowledge, this is the widest analysis ever conducted on the topic.
}


\maketitle

\section{Introduction}
\label{sec:introduction}
Cloud computing has changed the way we conceive and operate software. Nowadays, we can easily rent the (virtual) resources needed to run applications on a \textit{pay-per-use} basis~\cite{armbrust2010view}. A Virtual Machine (VM), a remote environment that works as a stand-alone computer with its own CPU, memory, network interface, and storage, can be accessed through proper APIs and exploited to run very diverse software systems.
While in the last years cloud providers started offering higher-level services such as container-as-a-service~\cite{caas_what_is} or function-as-a-service~\cite{serverless_what_is}, VMs remain key for many users, and the aforementioned services are often based on VMs themselves.
Because the behavior of provisioned VMs impacts the quality of the software running on it~\cite{DBLP:journals/jss/TianTL20}, be it a hosted application or a higher-level service (e.g., a container engine), it is crucial to understand and predict the performance of cloud platforms.

Such platforms foster scalability, flexibility, lightweight infrastructures, and lower costs, but they do not guarantee any (foreseen) performance per se~\cite{performance_challenges_gesvindr,qos_cloud_computing_Ardagna}. 
Moreover, the cloud hides a number of factors to the user: how hardware resources are managed and shared among users (multi-tenancy), which virtualization system is employed underneath, how and when applications are migrated to other computers, and how isolated they are. These are only a few examples of what is not transparent to the user and that forces us to consider VMs as \textit{black-box}es. 
As a consequence, cloud performance exhibits inherent variability that is challenging to comprehend and accurately predict \cite{patterns_leitner}.

Most cloud vendors do not provide a definition of ``performance'', neither do they supply a clear service level agreement (SLA) for their VMs: to the best of our knowledge, up-time ---which measures service availability--- is often the only collected and publicly available datum~\cite{azure_status,aws_status}. 
Some providers supply a maximum and minimum value for specific metrics: for example, they specify the number of input/output operations per second for disk storage (IOPS), to measure the number of reads from and writes to non-contiguous storage locations, and the throughput (per second), to state how fast the storage can read/write data~\cite{azure_disks,aws_disks,gcp_disks}. However, these metrics and data do not consider performance variability and do not help users gain insights on the expected performance.
Thus, assessing the qualities of service supplied by the VMs offered by the different cloud providers is a needed step towards creating performance models~\cite{qos_cloud_computing_Ardagna} and deriving quality indicators~\cite{cloudcmp_li}. Benchmarking has been widely used in this context: a significant number of works has already used application benchmarks~\cite{cloud_benchmark_suite_scheuner}, micro benchmarks~\cite{microbenchmarking_laaber}, and even cloud-specific benchmarks~\cite{ec2_performance_analysis_dejun} to evaluate the performance of VMs, but getting reliable results is not trivial and test-flawed methodologies may lead to incorrect conclusions~\cite{repeatable_experiments_abedi}.

This paper presents an in-depth analysis of VM performance that considers 16 types of VMs from 4 different cloud providers, namely: Amazon Web Services\footnote{https://aws.amazon.com/}, Microsoft Azure\footnote{https://azure.microsoft.com/}, Google Cloud Platform\footnote{https://cloud.google.com/gcp/}, and EGI\footnote{https://www.egi.eu} (European Grid Infrastructure). 
The first three are public and commercial cloud platforms, whereas EGI is a European federation of cloud providers that supports researchers and multinational projects. Besides studying well-known solutions, we also wanted to consider what is freely available to European researchers and understand possible differences.
While most of the studies~\cite{patterns_leitner} available in the literature are limited to few benchmarks and metrics (e.g., latency), our work exploits \benchSuite, a novel benchmark suite that we developed, that allows for systematic and recurrent retrieval of $28$ different metrics from $10$ different benchmarks.
\benchSuite probes the cloud platform following Randomized Multiple Trials (RMT) \cite{repeatable_experiments_abedi}, a well-known methodology for conducting repeatable experiments and enabling fair comparisons in the cloud, where the context may change frequently. 

%to provide an updated and comprehensive understanding of the variability of VMs. 

After collecting data, we synthesized an indicator, called \fullindicator (\indicator), to help quantify the variability of specific VMs or vendors' infrastructures in terms of breadth, frequency, and speed of performance fluctuations. 
The use of \indicator simplifies the understanding of the performance provided by cloud platforms task and allows one to capture the main trends of VM variability. For example, our results show that AWS and EGI are the more stable providers, while Azure and EGI are, on average, less reliable than AWS and GCP.

We then analyzed the collected data and metrics using four different dimensions to better characterize variability: i) \textit{resources} which studies how single resources (e.g., CPU, memory) are affected by variability, ii) \textit{isolation} that analyses whether cloud providers are able to isolate resources and their performance fluctuations, iii) \textit{time} that focuses on detecting recurring patterns and cyclic behaviors, and iv) \textit{cost} which analyzes whether renting more expensive VMs means obtaining better and higher performance.
Executed benchmarks allowed us to probe systematically cloud platforms and analyze the behavior of different solutions over a significant amount of time, to identify accurate variability distributions~\cite{statistics_he}. 

\nw{Finally, we developed multiple predictive models based on Machine Learning (ML) that allow for forecasting future performance and anticipate recurring patterns. In particular, we trained multiple time-series forecasting models to predict the performance of single resources.
Furthermore, we leveraged multivariate models, such as logistic regression, to predict various temporal factors, including the time of day, the day of the week, and weekends, based on a given set of measurements. These predictions, on average, exhibited an accuracy that was approximately twice as high as that achieved by a random model. This suggests that, despite the presence of noisy data, our findings confirm the existence of recurring patterns, providing valuable insights into the underlying performance dynamics.}

\nw{In summary, the contributions of this paper are the following: \textit{(i)} a new tool, called \benchSuite, for systematically retrieving and analyzing performance data of running VMs, \textit{(ii)} an indicator called \indicator (\fullindicator) for measuring performance variability of VMs, \textit{(iii)} a comprehensive multi-dimensional analysis of collected data, and \textit{(iv)} a set of ML-based predictive models to forecast VM performance.}

% \begin{itemize}[itemsep=0pt,parsep=0pt,topsep=0pt,partopsep=0pt]
%     \item a new tool, called \benchSuite, for systematically retrieving and analyzing performance data of running VMs;
%     \item an indicator, called \indicator, for measuring performance variability of VMs;
%     \item a comprehensive multi-dimensional analysis of collected data;
%     \item a set of ML-based predictive models to forecast VM performance.
% \end{itemize}

The rest of this paper is organized as follows.
Section \ref{sec-methodology} presents the methodology of our work. 
Section~\ref{sec-vi} introduces \fullindicator,
Section~\ref{sec-analysis} illustrates our multi-dimensional analysis, while 
Section~\ref{sec-ml} describes the ML-based predictive models.
Section~\ref{sec-discussion} discusses obtained results and answers to the research questions.
Section~\ref{sec-related} surveys the state of the art, and Section~\ref{sec-conclusions} concludes the paper.

\section{Methodology}
\label{sec-methodology}

\newcommand\rqvi{$\bm{RQ_1}$\xspace}
\newcommand\rqvitext{\textit{How can one measure the performance variability of VMs?}\xspace}

\newcommand\rqres{$\bm{RQ_2}$\xspace}
\newcommand\rqzero{\textit{Does the performance of commonly used VMs vary w.r.t. the different resources?}\xspace}

\newcommand\rqiso{$\bm{RQ_3}$\xspace}
\newcommand\rqthree{\textit{Are the resources provisioned to VMs managed in isolation?}\xspace}

\newcommand\rqtime{$\bm{RQ_4}$\xspace}
\newcommand\rqone{\textit{Is there a relationship between offered performance and time of the day or day of the week?}\xspace}

\newcommand\rqcost{$\bm{RQ_5}$\xspace}
\newcommand\rqfour{\textit{Are the performance of similar VMs offered by different providers comparable and proportionate to their \textit{cost}?}\xspace}

%
%-------------------------------------------------------------

The need for a shared, comprehensive knowledge base on VM performance variability motivated us to conduct an empirical assessment aimed to create an open, wide, and accessible dataset and to provide relevant insights on this phenomenon. Our work aims to answer the following research questions:

\nw{\textbf{\rqvi}: \rqvitext The performance of an ideal VM should not vary over time, and the provider should work on minimizing possible fluctuations. However, as indicated in the literature, VMs frequently exhibit unpredictable performance patterns. This question aims to identify an indicator to measure performance variability of VMs.}

\textbf{\rqres}: \rqzero Besides confirming previous results, i.e., that overall stability is not attainable, this question studies variability w.r.t the \textit{resources} behind a VM (i.e., CPU, memory, network, disk) and dis-aggregates how they contribute to variability. 

\textbf{\rqiso}: \rqthree This question addresses variability \textit{isolation}. Resources cooperate to execute applications, but we do not know whether cloud providers are able to isolate the variability on a resource with no side-effects on others in the short term.

\textbf{\rqtime}: \rqone This question investigates how the \textit{time} dimension could impact variability. If variability were governed by predictable periods, one could cope with them and adopt proper countermeasures. 

% \newcommand\rqtwo{\textit{Are VM types, sizes, and offered performance correlated? How does performance vary among different VMs of the same provider?}\xspace}

% \textbf{\rqcost}: \rqtwo This question studies variability w.r.t. the \textit{type/size} of the VM. It tries to understand whether there is a relationship between sizes, types, and performance variations since we already know that ``identical'' VM instances (same configuration, region, OS, and disk type), which should perform the same, vary significantly~\cite{patterns_leitner,schad_runtime_measurements,cerotti_flexible_cpu_provisioning}. 

\textbf{\rqcost}: \rqfour This question addresses the variability among \textit{similar} VMs, that is, VMs that offer the same characteristics to the user, but that might exploit different hardware infrastructures. If one pays a similar amount of money ---with the exception of EGI that is free--- and rents similar VMs, how different measured performance is.

\nw{To answer these questions we proceeded in three steps. First, we created a benchmark suite, called \benchSuite, to collect data on $16$ VM types offered by $4$ cloud providers, employing $10$ different benchmarks for a total of $28$ performance metrics. Second, we defined a variability indicator (\indicator) and analyzed the collected data using descriptive statistics. Third, we created a set of ML models to understand whether VM performance are predictable, that is, their variability is not random but shows recurrent trends and patterns.}

In the remaining of this section, we describe in details our benchmark suite and the metrics it supports, the selected VMs that we used in our evaluation, and the data collection process.

\subsection{VMBS}
\label{subsec-benchmark}

\begin{table}[t]
\renewcommand\arraystretch{1.5}
\setlength{\tabcolsep}{4.5pt}
\footnotesize\centering
\begin{tabular}[t]{@{}r|lll@{}}
\multicolumn{1}{c}{\textbf{Benchmark}} & \multicolumn{1}{c}{\textbf{Metric}} &
\multicolumn{1}{c}{\textbf{Meaning}} & \multicolumn{1}{c}{\textbf{Unit}} \\ \hline
\multicolumn{4}{c}{\textit{CPU}} \\ \hline
\textit{Sysbench} & CPU EVENTS & events (e) per second & e/s \\
 & CPU LAT & avg latency & ms \\
 & CPU TH LAT & threads avg latency & ms \\ \cline{1-2}\cline{3-4}
\textit{Nench} & CPU SHA256 & SHA256 execution & s \\
 & CPU BZIP2 & bzip2 execution & s \\
 & CPU AES & AES execution & s \\ \cline{1-2}\cline{3-4}
\textit{CPUBench} & CPU DUR & mean duration & s \\ \hline
\multicolumn{4}{c}{\textit{Network}} \\
\hline
 \textit{Nench} & NET 1 & DL - Cachefly CDN & MiB/s \\
 & NET 2 & DL - Leaseweb (NL) & MiB/s \\
 & NET 3 & DL - Softlayer DAL (US) & MiB/s \\
 & NET 4 & DL - Online.net (FR) & MiB/s \\
 & NET 5 & DL - OVH BHS (CA) & MiB/s \\ \cline{1-2}\cline{3-4}
\textit{Download} & NETB 1 & DL - url 1 (1 GB) & MiB/s \\
 & NETB 2 & DL - url 2 (100 MB) & MiB/s \\ 
\hline
\end{tabular}
\renewcommand\arraystretch{1.41}
\hspace{1cm} % or your desired distance
\begin{tabular}[t]{@{}r|lll@{}}
\multicolumn{1}{c}{\textbf{Benchmark}} & \multicolumn{1}{c}{\textbf{Metric}} &
\multicolumn{1}{c}{\textbf{Meaning}} & \multicolumn{1}{c}{\textbf{Unit}} \\ \hline
\multicolumn{4}{c}{\textit{Memory}} \\ \hline
\textit{Sysbench} & MEM SPEED & speed & MiB/s \\
 & MEM LAT & avg latency & ms \\ \hline

 \multicolumn{4}{c}{\textit{Disk}} \\ \hline
\textit{Sysbench} & DISK FILE R & file op - read & reads/s \\
 & DISK FILE W & file op - write & writes/s \\
 & DISK FILE F & file op - fsync & fsyncs/s \\
 & DISK THR R & throughput - read & MiB/s \\
 & DISK THR W & throughput - write & MiB/s \\
 & DISK LAT & avg latency & ms \\ \cline{1-2}\cline{3-4}
\textit{Nench} & DISK SEEK & ioping - avg seek rate & $\mu$s \\
 & DISK SEQ R & ioping - seq. read speed & MiB/s \\
 & DISK SEQ W & dd - avg seq. write speed & MiB/s \\ \cline{1-2}\cline{3-4}
\textit{DDBench} & DISKB LAT & dd - s. blk (latency) & MB/s \\
 & DISKB THR & dd - l. blk (throughput) & MB/s \\ \hline
\multicolumn{4}{c}{\textit{App}} \\ \hline
\textit{WebBench} & APPB & requests per second & req/s \\ \hline
\end{tabular}
\caption{Studied metrics.}
\label{tab:measured_values}
\end{table}

Traditional benchmarks focus on computing few high-level indicators such as time to complete a task, throughput, and response time. These application-oriented benchmarks may be useful to get information about the performance of a specific application, but they lack means to understand the performance variability of single resources in isolation (e.g. CPU, memory, or network). 

In this paper, we introduce \benchSuite\footnote{Source code available at \url{https://github.com/deib-polimi/VMBS-tool}.} (VM Benchmark Suite), which in contrast is created to probe the different resources provided by VMs: namely CPU, memory, disk, and network.
\benchSuite also follows the eight methodological principles proposed to correctly measure and report performance-related studies in the cloud~\cite{papadopoulos_principles_reproducible_evaluation}. 

\paragraph{Metrics}
\benchSuite aims to help measure both hardware and application performance. The suite is intended to treat CPU, memory, disk, and network independently. We selected and created light-weight benchmarks with a small number of dependencies to minimize unnecessary installation overhead. They also aimed to probe VMs and all the resources of interest thoroughly, while trying to minimize repetitions and execution time. 
\benchSuite bundles three resource-specific benchmarks (\textit{CPUBench}, \textit{DDBench}, and \textit{DownloadBench}) and one application-specific benchmark we developed (\textit{WebBench}), and two existing suites (\textit{Sysbench\footnote{\url{https://github.com/akopytov/sysbench}}} and \textit{Nench\footnote{\url{https://github.com/n-st/nench}}}). We developed custom benchmarks to better probe the different VMs.
\textit{Sysbench} and \textit{Nench} are two popular suites created to assess server performance, and bundles CPU, memory, disk, and network benchmarks. Similarly, \textit{WebBench} is an exemplar benchmark application that exploits multiple resources sequentially and/or concurrently. It mimics a client-server application and comprises 
a web server (\textit{gunicorn}\footnote{\url{https://gunicorn.org}}) and a load generator (\textit{wrk}\footnote{\url{https://github.com/wg/wrk}}). 

These benchmarks allowed us to measure 28 different metrics as reported in Table~\ref{tab:measured_values}. Such an in-depth, multi-faceted analysis of VM variability, along with a publicly available companion dataset, aims to help researchers and practitioners better understand the complex dynamics of cloud environments and motivate new research efforts.
We selected diverse benchmarks for the same resource to get more detailed results. We used three benchmarks to assess CPU performance.
\textit{SysBench} is based on the computation of prime numbers and measures: (i) the number of operations performed within a given time window (CPU EVENTS), (ii) the average latency (CPU LAT), and (iii) the average thread latency (CPU TH LAT), given it exploits highly concurrent threading. \textit{Nench} measures CPU performance with 3 algorithms: the time to (i) hash, with algorithm SHA 256 (CPU SHA256), (ii) compress, with algorithm bzip2 (CPU BZIP2), and (iii) encrypt, with algorithm AES (CPU AES), a 500 MB file. Our \textit{CPUBench} (CPU DUR) measures the time required to compute a trigonometric identity multiple timesto saturate the CPU. 

As for memory, we exploited \textit{SysBench} to collect the sequential write speed (MEM SPEED) and latency (MEM LAT). We only used one benchmark since memory is easy to probe and we wanted to save time.
We deeply tested disk performance with three benchmarks. \textit{SysBench} allowed us to measure the random read (DISK FILE R), write (DISK FILE W), and fsync (DISK FILE F) speed on a 1GB file. We also measured the file read and write throughput (DISK THR R and DISK THR W) along with the latency (DISK LAT). \textit{Nench} allowed us to assess disk seek rate (DISK SEEK), sequential read (DISK SEQ R), and write speed (DISK SEQ W). 
Our \textit{DDBench} measured disk latency, by writing multiple times a small-sized block (DISKB LAT), and throughput, by writing a large block once (DISKB THR). \textit{DDBench} takes advantage of the well-known command \texttt{dd} used to monitor the sequential writing performance of a disk device on a Unix-like system. 
Network speed is measured by downloading a file from multiple network sources, both through \textit{Nench} (NET 1-5) and \textit{DownloadBench} (NETB 1-2). We used multiple providers to better measure network speed on VM's side, and discarded providers that showed non-negligible slowdowns in their connectivity. 
We only considered one application benchmark, \textit{WebBench} (APPB), to widen our analysis, even if the focus was on isolated resources, and also to understand the comprehensive behavior of the different VMs.

\begin{figure}[t]
    \centering
\includegraphics[width=0.50\linewidth]{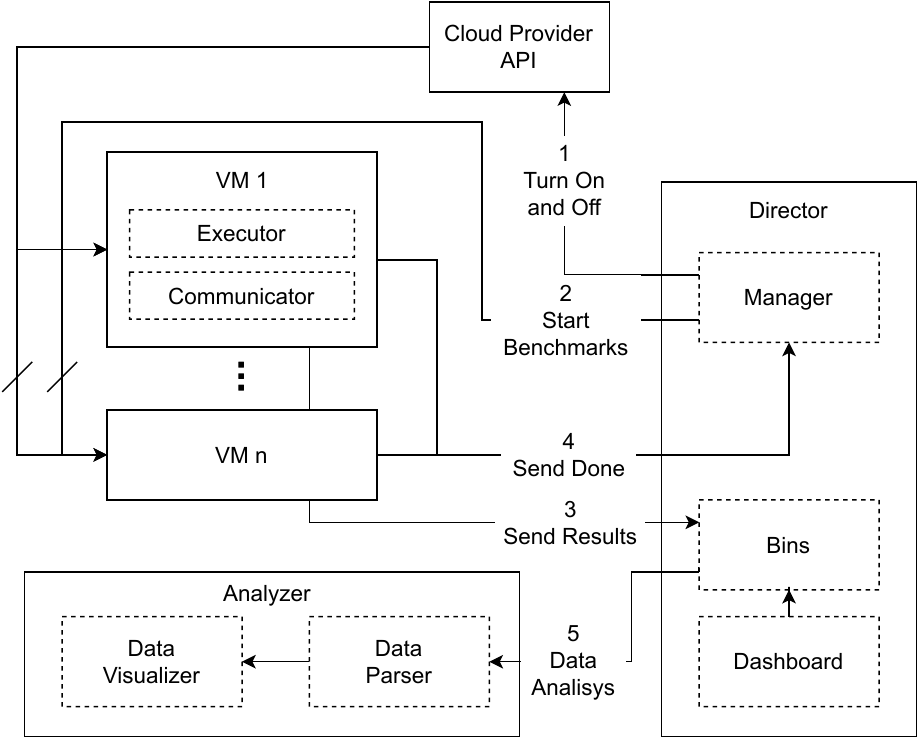}
    \caption{\benchSuite architecture.}
    \label{fig:implementation}
\end{figure}

\paragraph{Execution}
To collect data from different VMs and extract the aforementioned metrics, we implemented \benchSuite following the architecture presented in Figure~\ref{fig:implementation}.
\benchSuite automatically run a \textit{round} of experiments every hour. Each benchmark in a round is executed multiple times, called \textit{trial}s. The executions exploits the Randomized Multiple Trials (RMT) methodology, a simpler version of the Randomized Multiple Interleaved Trials (RMIT) approach~\cite{repeatable_experiments_abedi}. The RMT randomly reorders the execution sequence at each trial to mix the benchmark execution order and get more precise and unbiased results.

To minimize time and consumed resources, \benchSuite creates the VMs once: it then turns them on and off, but did not delete them for the whole duration of the experimentation (configurable by the users). The initial setup is designed to be minimal and only comprised the installation of our tool, the benchmarks, and all the dependencies required to execute them.

The experimental execution adheres to the following process.
Component \textit{Manager} periodically starts a round by turning on allocated instances and launching all benchmarks on every VM through \textit{Executor}. When the round completes and all results are available, \textit{Communicator} sends them to \textit{Bins} (a minimal, lightweight database that contains a \textit{bin} for each VM) and issues a done signal to \textit{Manager}, which turns the VMs off. \textit{Dashboard} shows the information about benchmark executions and VM status for monitoring purposes. The raw retrieved outputs are then processed by \textit{Data Parser} to clean them and extract the values obtained from each VM and for each metric. Parsed data are then fed to \textit{Data Visualizer}.

\subsection{VM Selection}
\label{subsec-usedvms}

This work targets VMs hosted in Europe, based on Ubuntu Linux, and provided by Amazon Web Services (AWS), Microsoft Azure, Google Cloud Platform (GCP), and the European Grid Infrastructure (EGI). 
Table~\ref{tab:instances} shows the VMs we used and groups them in four classes.Each class includes VMs with the same amount of CPUs and memory (RAM); the underlying non-virtualized hardware architecture may differ. Although the VMs in a class are not identical, the goal of this classification is to ease the comparison among similar VMs supplied by different providers. For each class (column $CL$), we show the number of VM instances we used for each provider, the number of virtual CPUs for each VM (column $vC$), the amount of available memory, its provider and code name, and its cost.\footnote{Shown prices refer to November 2022.} Selected configurations are popular and general-purpose ones, and provide a balanced amount of computing, memory, and networking resources.
These VMs are suited to run small and mid-sized applications, such as, micro-services, DBMS, web servers, code repositories. The main differences among the four classes are the number of vCPUs and the amount of memory.

\begin{table}[t]
\renewcommand\arraystretch{1.1}
\centering
\begin{tabular}{c|cccccc}
\textbf{CL} & \# & 
\textbf{vC} & \textbf{RAM} & 
\textbf{Provider} & \textbf{Name} & \textbf{Cost}  \\ \hline
\multirow{4}{*}{\textit{C1}} & \multirow{4}{*}{2} & \multirow{4}{*}{2} & \multirow{4}{*}{4 GiB} & AWS & a1.large & \$0,0582/h  \\ 
 &  &  &  & Azure & A2 v2 & \$0,0870/h   \\ 
 &  &  &  & GCP & E2-T1 & \$0,0713/h \\  
 &  &  &  & EGI & T1 & free  \\ \hline
\multirow{4}{*}{\textit{C2}} & \multirow{4}{*}{1} & \multirow{4}{*}{4} & \multirow{4}{*}{8 GiB} & AWS & a1.xlarge & \$0,1164/h   \\ 
 &  &  &  & Azure & A4 v2 & \$0,1830/h   \\
 &  &  &  & GCP & E2-T2 & \$0,1425/h \\ 
 &  &  &  & EGI & T2 & free \\ \hline
\multirow{4}{*}{\textit{C3}} & \multirow{4}{*}{2} & \multirow{4}{*}{2} & \multirow{4}{*}{8 GiB} & AWS & m5.large & \$0,1150/h \\ 
 &  &  &  & Azure & B2MS & \$0,0960/h  \\
 &  &  &  & GCP & N1-T1 & \$0,1250/h  \\ 
 &  &  &  & EGI & T3 & free   \\ \hline
\multirow{4}{*}{\textit{C4}} & \multirow{4}{*}{1} & \multirow{4}{*}{4} & \multirow{4}{*}{16 GiB} & AWS & m5.xlarge & \$0,2300/h  \\ 
 &  &  &  & Azure & B4MS & \$0,1920/h \\ 
 &  &  &  & GCP & N1-T2 & \$0,2501/h \\ 
 &  &  &  & EGI & T4 & free \\ \hline
\end{tabular}
\caption{Selected VMs.}
\label{tab:instances}
\end{table}

AWS a1 VMs are cost-saving machines designed to scale-out workloads, while m5 ones balance computing, memory, and network resources. Azure A2 and A4 machines are affordable and general purpose VMs; B ones instead are supposed to serve workloads that typically need low to moderate CPU power, but sometimes need to burst to significantly higher performance when the workload increases. As for GCP, we selected E2 and N1 VMs: the former ones favor cost optimization rather than performance; the latter are the first generation, general-purpose VMs offered by GCP. While we only used their CPUs, we selected N1 VMs since they also offer GPUs and can thus be more versatile and appropriate for very diverse contexts. Finally, EGI provides a set of predefined VM sizes, but they also allow one to exploit custom VMs based on specific needs. We created special-purpose VMs that mimicked the characteristics of the others.
Several independent providers contribute to the federation; our instances were hosted at the INFN (National Institute of Nuclear Physics) center in Padua (Italy).

Cloud providers also offer different disk types. The selected ones are those recommended for most of the workloads. In particular: the AWS General Purpose (GP2) SSD (Solid State Drive) for AWS, the Standard Azure SSD for Azure, and the default boot-persistent SSD disk for GCP. Azure also augments its VMs with a temporary disk that provides short-term storage for applications and processes and it is intended to only store data. EGI does not share any information about available disks. All run benchmarks only used the local boot disk.
We exploited the Command Line Interface (CLI) provided by AWS, Azure, and GCP to interact with their VMs. Since EGI is deployed onto OpenStack, we managed its VMs through its API.

\subsection{Data Collection}

The analysis presented in this paper is based on one-month experiments carried out in 2020. The measurements were collected in April, May, July/August and September/October, 2020, for AWS, Azure, GCP, and EGI, respectively. Our dataset contains more than 1.5 million data points.\footnote{The dataset we collected is available from:~\url{https://doi.org/10.5281/zenodo.8014668}, and all graphs and extracted data from:~\url{https://github.com/deib-polimi/VMBS-tool-Analyzer/}.}

\begin{table}[t]
\renewcommand\arraystretch{1.1}
\centering
%\begin{adjustbox}{width=0.49\textwidth}
\begin{tabular}{r|ccccc}
\textbf{Provider} & \textbf{Rounds} & \textbf{Trials} & \textbf{Measures} & \textbf{ErrBe} & \textbf{ErrPr} \\ \hline
\textit{AWS} & 4,866 & 165,437 & 442,663 & 0.004\% & 0\% \\ \hline
\textit{Azure} & 3,744 & 127,283 & 351,780 & 0.010\% & 9.935\% \\ \hline
\textit{GCP} & 4,721 & 160,514 & 443,734 & 0\% & 0\% \\ \hline
\textit{EGI} & 4,344 & 135,258 & 395,823 & 8.421\% & 0\% \\ \hline
\end{tabular}
%\end{adjustbox}
\caption{Experiments, repetitions, and errors.}
\label{tab:execution_info}
\end{table}

We executed a round every hour for (a bit longer than) a month. Table~\ref{tab:execution_info} shows the number of times \benchSuite was executed on the six different VMs of each provider (Rounds). 
We executed each benchmark different times (from 10 to 3 times, see below) and also used different configurations (e.g., \textit{Download Bench} was executed five times with a 1 GB file and five times with a 100 MB file). During each round, we executed \textit{DDBench} and \textit{DownloadBench} 10 times, 5 times for each considered configuration, \textit{CPUBench} 5 times, and \textit{Sysbench}, \textit{Nench}, and \textit{WebBench} 3 times each, for a total of 34 trials. These numbers are a compromise between total execution time and statistical validity of obtained results. Column Trials represents the total number of executed trials: the numbers reported in the table are lower than expected since some trials failed (column \textit{ErrBe}), and we only report successful ones. Column \textit{Measures} refers to the number of retrieved values (for all considered metrics). For example, Sysbench returns 11 values and is repeated three times, and thus returns 33 values at each round. Also in this case, some benchmarks did not return all values in each trial, and thus the total is lower than the theoretical value. These missed values are not reported in column ErrBe because we did not consider as erroneous benchmarks that ran without problems but did not return some values. 
For example, DISK SEEK and DISK SEQ R did not produce any value on ARM-based VMs because used libraries do not work with those machines. Note also that some attempts to restart VMs on Azure were unsuccessful (column \textit{ErrPr}). Similarly, only EGI gave problems executing (network) benchmarks. 

The mean duration ---over a month--- of each round (full execution of \benchSuite) on the different VMs was different among providers even if we used similar VMs: some $10$ minutes with AWS, $20$ minutes with Azure, $14$ minutes with GCP, and $22$ minutes with EGI. 

\begin{figure}[b]
\centering
\includegraphics[width=0.52\linewidth]{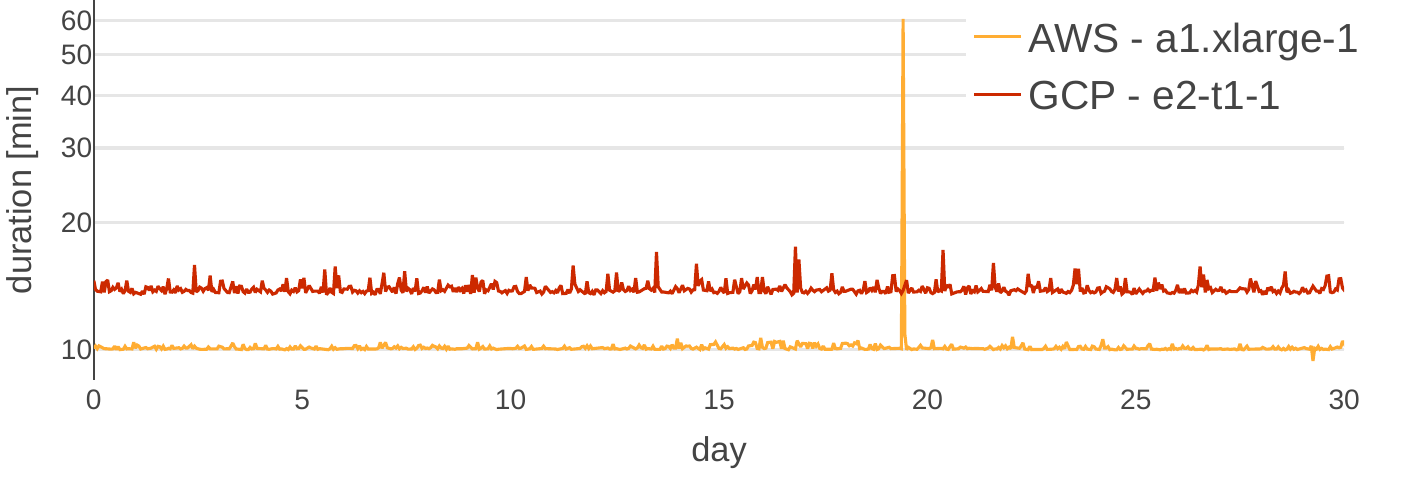}
\caption{\benchSuite on AWS a1.xlarge-1 and on GCP E2-T1-1.}
\label{fig:duration}
\end{figure}

To exemplify the fluctuating execution times of \benchSuite on the different providers, Figure~\ref{fig:duration} compares the execution of \benchSuite on AWS a1.xlarge and GCP E2-T1. The y-axis shows the duration in minutes while the x-axis reports the time at which the toolkit was executed. Other VMs followed similar behaviors. The execution times on AWS remained stable during the whole month, with a standard deviation ---relative to the mean execution time--- lower than $1.80$ minutes for all machines. They only presented a spike at the same time for all machines.While with more ups and downs, also the execution times on GCP were stable, with a standard deviation always lower than $0.48$ minutes for all machines.

\begin{figure}[t]
	\centering
	\includegraphics[width=0.52\linewidth]{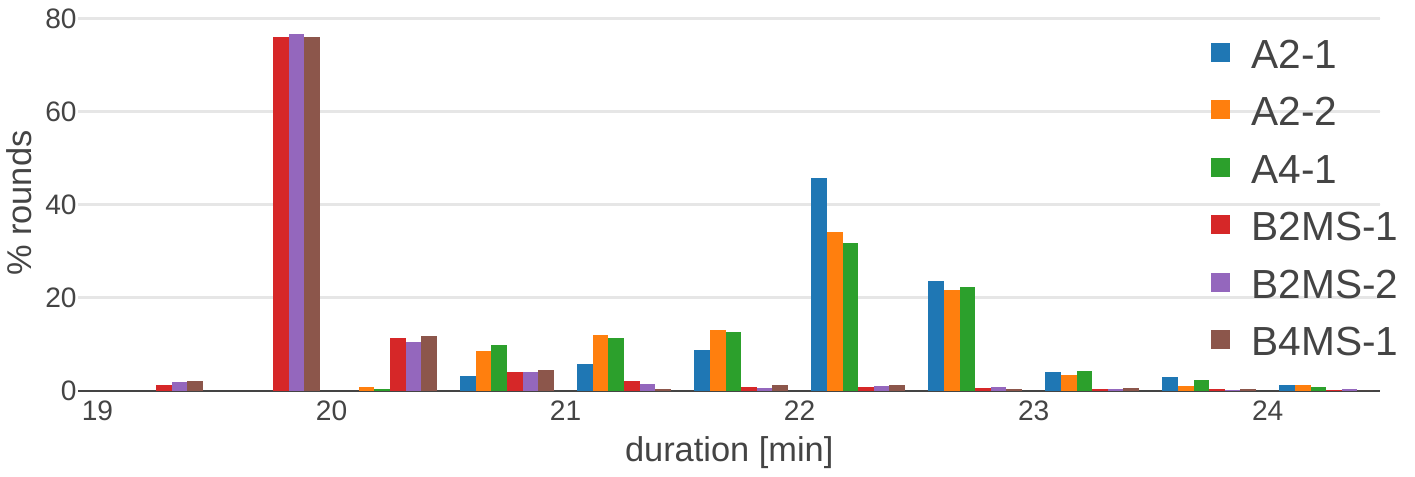}
	\caption{Distribution of execution times on Azure (marginal values are not shown).}
	\label{fig:duration_distribution_azure}
\end{figure}

In contrast, the duration on Azure is more variable (Figure~\ref{fig:duration_distribution_azure}): the graph shows multiple peaks at different durations of different magnitudes. The duration of a round on B*MS machines was around 20 minutes in most of the cases (75\%), while on A* machines it was more variable: between 20.5 to 23.5 minutes. The interval between the 19\textsuperscript{th} and 25\textsuperscript{th} of May presented a slowdown common to all the machines. The minimum standard deviation is $1.05$ and the maximum is $7.72$ minutes for the different VMs. 
Due to the errors that emerged during the execution of network benchmarks on EGI, we decreased the size of one of the downloaded file used to measure network speed. The execution time with EGI is variable and there is a sharp reduction at the middle of the observation period. This is due to the fact that the VMs were not able to reach the online resource anymore, thus the download benchmarks were partially skipped, and the total execution time decreased.

\section{\fullindicator}
\label{sec-vi}

To answer \rqvi and to have an initial coarse-grained view of the performance variability of VM, we synthesized a \textit{Variability Indicator} (\indicator).
\indicator uses historical data to quantify variability using three dimensions: i) \textit{breadth}, that is, how big the observed changes in the values were or how big the differences in the quality of the service were, ii) \textit{dispersion}, that is, how scattered collected values are, and
\textit{speed}, that is, how fast variations occurred.
These three dimensions are helpful to understand whether measured degradation is acceptable in the short-term.
Note that, \indicator quantifies how variable the performance of the VM was in a given, user defined, time window. On the contrary, it does not predict the future behavior of the VM (we present a set of complementary predictive models in Section~\ref{sec-methodology}).

For each metric $m$ listed in Table~\ref{tab:measured_values}, a vector $X_m$ contains the sequence of all collected values: one entry per hour since we assumed each metric to be constant for the duration of a single round.
The vector is then used to create a variation vector $V_m$, where each element $v_m$ is the difference (as percentage) between the corresponding $x_m$ and $\mu_m$ (the average of all values in $X_m$), normalized by $\mu_m$: 

\begin{equation} \label{eq:div_vector}
v_m = \frac{x_m-\mu_m}{\mu_m}\cdot 100
\end{equation}

\noindent $V_m$ is then used to create a change vector $C_m$. For some metrics where higher is better (HIB), such as CPU EVENTS or MEM SPEED, $C_m$ contains $v_m$ if it is lower than a threshold $t$, and zero otherwise.
For other metrics where lower is better (LIB), such as CPU BZIP2 or DISK LAT, $C_m$ contains $v_m$ if it is greater than $t$ and zero otherwise.
Note that $t$ is used to filter out small variations: for example, $t = 0.1$ would only consider variations greater than $10\%$ (with respect to the mean value). The result is a vector with values that represent changes or zeros. 

%\begin{equation} \label{eq:degradation_vector_hib}
%C_m = 
%\begin{cases}
%    0,& \text{if $v_m$ $\geq$ t}\\
%    v_m,& \text{if $v_m$ $<$ t}
%\end{cases},\quad \forall v_m \in V_m
%\end{equation}
%
%\begin{equation} \label{eq:degradation_vector_lib}
%C_m = 
%\begin{cases}
%    0,& \text{if $v_m$ $\leq$ t}\\
%    v_m,& \text{if $v_m$ $>$ t}
%\end{cases},\quad \forall v_m \in V_m
%\end{equation}

Breadth $bdth_m$ is then computed as the module of the average of the values in $C_m$. Note that, given each $v_m$ is computed as difference with respect to the average $\mu_m$, this value can be interpreted as an estimation of both degradation and improvement over the period. A $bdth_m$ equal to $0$ would then mean a stable performance. 

%over a period $T$, where $T$ must be $\leq | D_m |$ and is equal to the number of samples of interest, is computed as the summation of the relative negative variations multiplied by their duration \luc{La frase precedente \`e corretta?}. Each term of the summation is the area of a trapezoid with basis $D_m(t)$ and $D_m(t-1)$ and with height $1$ (step), the sampling rate (in our case one hour) (Formula~\ref{eq:degradation}). The result of the summation is divided by the total observation time to obtain a relative degradation value, that is independent of the duration of the observation period. A negative degradation would mean an improvement, while a null degradation ($deg_m = 0$) would mean no degradation with respect to threshold $t$.

%\begin{equation} \label{eq:degradation}
%\begin{split}
%    T = | D_m^- | \\
%deg_m = \dfrac{\sum_{t=1}^{t=T} (D_m^-(t) %+ D_m^-(t-1)) \cdot s/2}{T \cdot s} \\
%      = \dfrac{\sum_{t=1}^{t=T} D_m^-(t) %+ D_m^-(t-1)}{2T}
%\end{split}
%\end{equation}

Dispersion $dis_m$ is computed as the relative standard deviation (RSD) of the data vector $X_m$, thus the standard deviation of $X_m$ divided by its mean value $\mu_v$ (Formula~\ref{eq:rsd}) and then expressed as percentage. $0$ means no variation.

\begin{equation} \label{eq:rsd}
dis_m = \dfrac{\sigma(X_m)}{\mu_m} \cdot 100
\end{equation}

\noindent Speed $speed_m$ is computed as the standard deviation of the gradient of collected samples divided by the mean value (expressed as percentage). The speed measures how fast the variation is, thus if the variation is null the speed should be zero. 

\begin{equation} \label{eq:speed}
speed_m = \dfrac{\sigma(\nabla(X_m))}{\mu_m} \cdot 100
\end{equation}

\noindent Our \textit{variability indicator} is then computed as a weighted sum of $bdth_m$, $dis_m$, and $speed_m$ (Equation~\ref{eq:pqi}). Therefore, the \indicator of an ideal VM should be $0$
.
\begin{equation} \label{eq:pqi}
\begin{split}
\indicator_m\ = bdth_m \cdot w_{b} + dis_m \cdot w_{d} + speed_m \cdot w_{s}
\end{split}
\end{equation}

\noindent where $w_b$, $w_d$, $w_s$ are the weights we want to adopt to value the three indicators: $w_b, w_d, w_s \in [0, 1]$ and $w_b + w_d + w_s = 1$.
The \indicator can be a simple and valid means to analyze the quality of provided VMs. The user may experiment with different values for $w_{b}$, $w_{d}$, and $w_{s}$ to carry out what-if analyses that emphasize different aspects, but the lower a \indicator is, the lower the performance variability of the VM is. % A kind of ideal \indicator should be equal to zero, to mean no variability. 

\begin{table*}[t]
	\renewcommand\arraystretch{1.3}
	\fontsize{8}{9.6}\selectfont
	\centering
	\begin{adjustbox}{width=1\textwidth}
		\begin{tabular}{l|ccc|c|ccc|c|ccc|c|ccc|c}
			\multicolumn{1}{c|}{} & bdth & dis & speed & \textbf{\indicator} & bdth & dis & speed & \textbf{\indicator} & bdth & dis & speed & \textbf{\indicator} & bdth & dis & speed & \textbf{\indicator} \\ \hline
			
			\multirow{2}{*}{\textbf{C1}} &  \multicolumn{4}{c|}{\textit{AWS - a1.large}} & \multicolumn{4}{c|}{\textit{Azure - A2-v2}} & \multicolumn{4}{c|}{\textit{GCP - E2-T1}} & \multicolumn{4}{c}{\textit{EGI - T1}}   \\
			& 2.50 & 6.80 & 3.99 & \cellcolor[HTML]{C0C0C0}\textbf{4.43 }
			& 16.33 & 41.74 & 25.71 & \cellcolor[HTML]{EFEFEF}\textbf{27.93 }
			& 3.51 & 8.97 & 6.02 & \textbf{6.17 }
			& 5.16 & 14.16 & 9.65 & \textbf{9.66 }
			\\ \hline
		      
			\multirow{2}{*}{\textbf{C2}} &  \multicolumn{4}{c|}{\textit{AWS - a1.xlarge}} & \multicolumn{4}{c|}{\textit{Azure - A4-v2}} & \multicolumn{4}{c|}{\textit{GCP - E2-T2}} & \multicolumn{4}{c}{\textit{EGI - T2}}   \\ 
			& 2.71 & 7.74 & 3.88 & \cellcolor[HTML]{C0C0C0}\textbf{4.78 }
			& 15.30 & 38.40 & 23.83 & \cellcolor[HTML]{EFEFEF}\textbf{25.84 }
			& 3.14 & 8.05 & 5.34 & \textbf{5.51 }
			& 6.92 & 20.24 & 13.46 & \textbf{13.54 }
			\\ \hline
			
			\multirow{2}{*}{\textbf{C3}} &  \multicolumn{4}{c|}{\textit{AWS - m5.large}} & \multicolumn{4}{c|}{\textit{Azure - B2MS}} & \multicolumn{4}{c|}{\textit{GCP - N1-T1}} & \multicolumn{4}{c}{\textit{EGI - T3}}   \\ 
			& 2.87 & 7.63 & 4.59 & \cellcolor[HTML]{C0C0C0}\textbf{5.03} 
			& 3.97 & 10.52 & 6.76 & \textbf{ 7.09 }
			& 3.85 & 9.66 & 6.66 &  \textbf{6.72 }
			& 6.75 & 20.81 & 14.11 & \cellcolor[HTML]{EFEFEF}\textbf{13.89} 
			\\ \hline
   
			\multirow{2}{*}{\textbf{C4}} &  \multicolumn{4}{c|}{\textit{AWS - m5.xlarge}} & \multicolumn{4}{c|}{\textit{Azure - B4MS}} & \multicolumn{4}{c|}{\textit{GCP - N1-T2}} & \multicolumn{4}{c}{\textit{EGI - T4}}   \\ 
			& 2.82 & 7.49 & 4.11 & \cellcolor[HTML]{C0C0C0} \textbf{4.81 }
			& 3.57 & 9.57 & 6.11 & \textbf{6.42 }
			& 3.47 & 8.89 & 6.07 & \textbf{6.15 }
			& 5.66 & 18.64 & 12.97 &\cellcolor[HTML]{EFEFEF} \textbf{12.42} \\
     \bottomrule
		\end{tabular}
	\end{adjustbox}
	\caption{Most significant \indicators (one metric for each considered resource) and mean values among all collected metrics (all values are percentages). Best \indicator are highlighted in dark gray, worst in light gray.}
	\label{tab:pqi}
\end{table*}

Table~\ref{tab:pqi} shows the most significant \indicators, one for each considered VM, aggregated on all the metrics. $VI$ is computed with $t=0$ and $w_{b} = w_{d} = w_{s} = 1/3$, that is, they all weigh the same since the three aspects are equally important. The value is averaged when we used multiple instances. While the \indicator can be calculated for every single metric, we report here the aggregated results to simplify the discussion and to provide a high-level result that takes into consideration all the analyzed metrics. For each class, the table uses a dark gray background to highlight the best \indicator, and a light gray one to identify the worst.

The results show that AWS VMs provide, on average, better and more stable performance compared to all the other providers.
In particular, for all the classes, AWS achieved the best score on $bdth$, $dis$, $speed$, and \indicator. In class C1 and C2, AWS obtained a \indicator of $4.43$ and $4.78$ that is, respectively, $28\%$ and $13\%$ lower than GCP, the second best provider.
EGI obtained significantly worse results with a \indicator that is $54\%$ and $64\%$ higher than AWS. Finally, Azure with VMs A2 and A4 obtained by far the highest \indicator, $81\%$ and $84\%$ worse than AWS. 

The results obtained for classes C3 and C4 confirm that AWS provides the most stable performance but with smaller differences among the providers. AWS machines obtained a \indicator of $5.03$ and $4.81$ that is, respectively $25\%$ and $21\%$ better than GCP, once again the second best provider. %As demonstrated by our multi-faceted analysis (e.g., $RQ_5$), 
The performance of larger Azure machines appear to be more stable than their smaller counterparts. 
They obtained results that are slightly worse than the ones of GCP with \indicators that are, respectively, only $5\%$ and $4\%$ higher. Finally, the worst performance for classes C3 and C4 was obtained by EGI with \indicators that are almost three times higher than the ones of AWS. 

%This confirms the data reported in Table~\ref{tab:mmm_all}, where among the 30 results shown for classes C3 and C4 (excluding CPR not reported for EGI), EGI obtained the worst performance in 11 metrics (more than any other provider).

\indicator allows one to estimate variability in a selected time frame for a large set of metrics. The results show that the \indicator is able to clearly capture the significant trends that emerged, in a much ``scattered'' way, from other analyses described in the following sections.
\indicator tells us that AWS provides the most stable performance across the cohort of analyzed cloud providers, closely followed by GCP, while Azure and EGI were not always as stable as the first two.

\section{Data Analysis}
\label{sec-analysis}
This section details our analysis on the collected data with the goal of answering the remaining research questions (\rqres-\rqcost).

\subsection{Resources}

\begin{table*}[t]
\renewcommand\arraystretch{1.3}
\setlength{\tabcolsep}{3.5pt}
\fontsize{8}{9.6}\selectfont
\centering
%\begin{adjustbox}{width=1\textwidth}
\begin{tabular}{r|ccc|ccc|ccc|ccc}
\textbf{} & \multicolumn{3}{c|}{\textbf{AWS}} & \multicolumn{3}{c|}{\textbf{Azure}} & \multicolumn{3}{c|}{\textbf{GCP}} & \multicolumn{3}{c}{\textbf{EGI}} \\ \hline
\textbf{Metric} & avg & min & max & avg & min & max & avg & min & max & avg & min & max \\ \hline
CPU EVENTS & 1.46\% & 0.92\% & 2.36\% & \cellcolor[HTML]{EFEFEF}4.76\% & 4.24\% & 5.75\% & 1.38\% & 1.02\% & 1.72\% & \cellcolor[HTML]{C0C0C0}1.19\% & 0.64\% & 1.69\% \\
CPU LAT & 3.16\% & 1.82\% & 5.14\% & \cellcolor[HTML]{EFEFEF}4.61\% & 3.93\% & 5.36\% & 2.18\% & 0.87\% & 2.97\% & \cellcolor[HTML]{C0C0C0}1.30\% & 0.31\% & 2.06\% \\
CPU TH LAT & 3.30\% & 2.32\% & 5.65\% & \cellcolor[HTML]{EFEFEF}9.74\% & 7.88\% & 13.17\% & \cellcolor[HTML]{C0C0C0}1.45\% & 1.20\% & 1.98\% & 2.60\% & 1.86\% & 4.77\% \\
CPU SHA256 & 1.75\% & 1.01\% & 2.75\% & \cellcolor[HTML]{EFEFEF}5.15\% & 4.67\% & 5.69\% & 2.14\% & 1.97\% & 2.32\% & \cellcolor[HTML]{C0C0C0}1.71\% & 1.23\% & 2.15\% \\
CPU BZIP2 & 1.92\% & 0.64\% & 3.76\% & \cellcolor[HTML]{EFEFEF}4.73\% & 3.75\% & 5.74\% & 2.68\% & 1.40\% & 3.18\% & \cellcolor[HTML]{C0C0C0}1.56\% & 1.01\% & 1.75\% \\
CPU AES & 4.00\% & 2.62\% & 8.94\% & \cellcolor[HTML]{EFEFEF}15.35\% & 13.98\% & 17.14\% & 3.35\% & 2.01\% & 4.35\% & \cellcolor[HTML]{C0C0C0}2.47\% & 1.94\% & 2.78\% \\
CPU DUR & 35.77\% & 25.71\% & 53.52\% & 22.66\% & 14.25\% & 31.43\% & \cellcolor[HTML]{EFEFEF}43.62\% & 28.51\% & 52.18\% & \cellcolor[HTML]{C0C0C0}21.22\% & 10.45\% & 31.44\% \\ \hline
MEM SPEED & 1.80\% & 0.83\% & 3.07\% & 7.93\% & 4.74\% & 9.10\% & \cellcolor[HTML]{C0C0C0}1.61\% & 1.05\% & 3.25\% & \cellcolor[HTML]{EFEFEF}9.27\% & 4.53\% & 13.38\% \\
MEM LAT & 2.88\% & 1.29\% & 5.65\% & 8.49\% & 4.84\% & 9.90\% & \cellcolor[HTML]{C0C0C0}1.95\% & 1.11\% & 4.08\% & \cellcolor[HTML]{EFEFEF}9.56\% & 3.78\% & 15.46\% \\ \hline
DISK FILE R & \cellcolor[HTML]{C0C0C0}5.38\% & 2.45\% & 9.01\% & \cellcolor[HTML]{EFEFEF}61.09\% & 4.65\% & 124.37\% & 9.37\% & 7.28\% & 10.58\% & 15.76\% & 1.97\% & 21.06\% \\
DISK FILE W & \cellcolor[HTML]{C0C0C0}5.38\% & 2.45\% & 9.01\% & \cellcolor[HTML]{EFEFEF}61.09\% & 4.65\% & 124.37\% & 9.37\% & 7.28\% & 10.58\% & 15.76\% & 1.97\% & 21.06\% \\
DISK FILE F & \cellcolor[HTML]{C0C0C0}5.38\% & 2.45\% & 9.01\% & \cellcolor[HTML]{EFEFEF}60.96\% & 4.65\% & 124.08\% & 9.36\% & 7.28\% & 10.57\% & 15.73\% & 1.97\% & 21.02\% \\
DISK THR R & \cellcolor[HTML]{C0C0C0}5.38\% & 2.45\% & 9.01\% & \cellcolor[HTML]{EFEFEF}61.09\% & 4.65\% & 124.37\% & 9.37\% & 7.28\% & 10.58\% & 15.76\% & 1.98\% & 21.06\% \\
DISK THR W & \cellcolor[HTML]{C0C0C0}5.38\% & 2.45\% & 9.01\% & \cellcolor[HTML]{EFEFEF}61.09\% & 4.65\% & 124.36\% & 9.37\% & 7.28\% & 10.58\% & 15.76\% & 1.98\% & 21.06\% \\
DISK LAT & \cellcolor[HTML]{C0C0C0}6.09\% & 2.10\% & 9.89\% & \cellcolor[HTML]{EFEFEF}36.61\% & 6.31\% & 69.55\% & 9.51\% & 7.44\% & 10.65\% & 18.99\% & 3.04\% & 28.94\% \\
DISK SEEK & \cellcolor[HTML]{C0C0C0}1.16\% & 1.09\% & 1.28\% & 19.01\% & 1.17\% & 36.13\% & 11.56\% & 9.80\% & 13.42\% & \cellcolor[HTML]{EFEFEF}98.03\% & 4.17\% & 154.95\% \\
DISK SEQ R & 0.63\% & 0.41\% & 1.03\% & \cellcolor[HTML]{EFEFEF}31.33\% & 0.32\% & 63.69\% & \cellcolor[HTML]{C0C0C0}0.27\% & 0.08\% & 0.57\% & 19.24\% & 2.21\% & 28.11\% \\
DISK SEQ W & \cellcolor[HTML]{C0C0C0}0.37\% & 0.15\% & 1.17\% & 0.89\% & 0.16\% & 2.67\% & 9.96\% & 7.38\% & 12.47\% & \cellcolor[HTML]{EFEFEF}11.49\% & 0.78\% & 16.10\% \\
DISKB LAT & 13.37\% & 11.96\% & 14.14\% & 87.34\% & 21.53\% & 197.30\% & \cellcolor[HTML]{C0C0C0}12.77\% & 10.10\% & 14.37\% & \cellcolor[HTML]{EFEFEF}102.98\% & 12.23\% & 141.24\% \\
DISKB THR & 5.96\% & 4.25\% & 7.49\% & \cellcolor[HTML]{C0C0C0}2.37\% & 0.38\% & 6.47\% & 13.04\% & 10.76\% & 16.79\% & \cellcolor[HTML]{EFEFEF}13.62\% & 3.62\% & 18.08\% \\ \hline
NET 1 & 8.99\% & 8.65\% & 9.36\% & \cellcolor[HTML]{EFEFEF}29.38\% & 21.79\% & 36.65\% & 7.04\% & 6.79\% & 7.25\% & \cellcolor[HTML]{C0C0C0}4.51\% & 4.40\% & 4.67\% \\
NET 2 & 16.13\% & 15.08\% & 16.97\% & \cellcolor[HTML]{EFEFEF}30.18\% & 27.94\% & 32.23\% & 27.41\% & 26.55\% & 28.59\% & \cellcolor[HTML]{C0C0C0}14.19\% & 13.52\% & 15.80\% \\
NET 3 & \cellcolor[HTML]{EFEFEF}24.39\% & 19.78\% & 26.58\% & 14.94\% & 6.99\% & 23.54\% & 10.28\% & 9.61\% & 10.83\% & \cellcolor[HTML]{C0C0C0}1.30\% & 1.19\% & 1.46\% \\
NET 4 & 10.46\% & 8.80\% & 11.75\% & \cellcolor[HTML]{EFEFEF}13.58\% & 10.64\% & 16.73\% & 9.51\% & 9.26\% & 9.89\% & \cellcolor[HTML]{C0C0C0}6.54\% & 6.03\% & 7.15\% \\
NET 5 & \cellcolor[HTML]{EFEFEF}10.81\% & 8.34\% & 12.32\% & 7.47\% & 6.66\% & 8.96\% & 6.98\% & 6.76\% & 7.53\% & \cellcolor[HTML]{C0C0C0}3.82\% & 3.49\% & 4.21\% \\
NETB 1 & \cellcolor[HTML]{C0C0C0}7.29\% & 6.49\% & 8.48\% & 25.88\% & 23.80\% & 28.23\% & 19.63\% & 18.51\% & 20.49\% & \cellcolor[HTML]{EFEFEF}78.82\%$^*$ & 71.01\%$^*$ & 86.77\%$^*$ \\
NETB 2 & 8.96\% & 8.49\% & 9.55\% & \cellcolor[HTML]{EFEFEF}19.04\% & 17.93\% & 20.18\% & 6.77\% & 6.63\% & 6.90\% & \cellcolor[HTML]{C0C0C0}2.85\% & 2.68\% & 3.06\% \\ \hline
APPB & 1.97\% & 0.80\% & 7.33\% & \cellcolor[HTML]{EFEFEF}4.89\% & 3.68\% & 6.39\% & \cellcolor[HTML]{C0C0C0}1.04\% & 0.70\% & 1.52\% & 1.74\% & 1.31\% & 2.30\% \\
\bottomrule
\end{tabular}
%\end{adjustbox}
\caption{Average, minimum, and maximum relative standard deviations among all VMs ($^*$ means that due to network problems, the size of downloaded file was smaller). Best results are highlighted in dark gray, worst in light gray.}
\label{tab:bench_results_all_rsd_avg}
\end{table*}

\begin{figure}[htbp]
	\centering
	\includegraphics[width=0.55\linewidth]{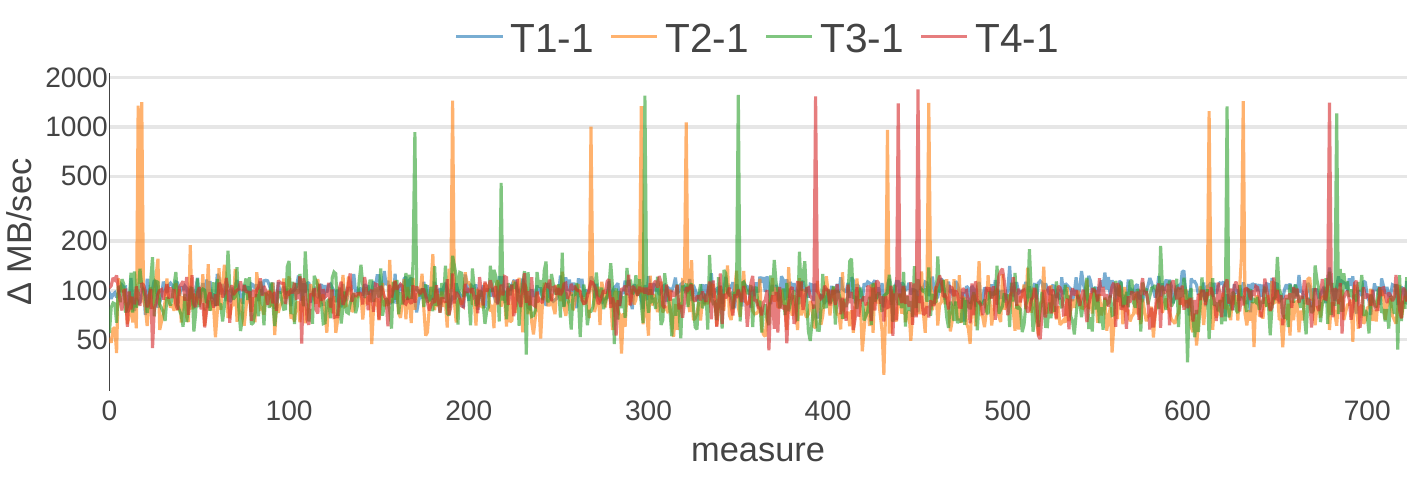}
	\caption{DISKB LAT on EGI VMs (measure identifies the ordinal of retrieved values).}
	\label{fig:egi_diskb_lat}
\end{figure}

\rqres concerns how single resources affect the variability of the performance of used VMs. %This question helped us frame the context and also set a baseline for the other questions.
%The answer is that sometimes the variability is limited and can be acceptable; in other cases, it is not. Moreover, if we consider a single provider, different resources are affected by different levels of variability.
As an example, Figure~\ref{fig:egi_diskb_lat} shows the results obtained for metric DISKB LAT with EGI VMs. If one assumed the mean value $\mu$ as reference, the chart shows that measured values varied without any evident pattern for the different VM types. They also varied between different VMs of the same type, but this is not exemplified in the figure.

Table~\ref{tab:bench_results_all_rsd_avg} introduces the relative standard deviations (RSD) of the measured valued for the $28$ metrics averaged on the $6$ VMs used for each provider.
For each metric (row), the table uses a dark gray background to highlight the best provider, and a light gray one to identify the worst. The values say that performance variation strictly depends on benchmarked resources.

As shown in the table, almost all the metrics are affected by non-trivial variations.
However, the results for each provider show that resources vary in different ways.
For example AWS is affected by an average variation of around $5\%$ on disk performance, while more than double ($12\%$) on networking.
Moreover, while AWS is consistently the most stable provider for what concern disk metrics, it is also the worst one in 2 tests over 7 related to networking.
Similarly, EGI obtains the lowest variability among all the providers on CPU- and network-related metrics, while the highest one with memory-related tests. 

AWS variability ranges from a minimum value of 0.15\% (DISK SEQ W) to a maximum value of 54\% (CPU DUR), Azure variability ranges from 0.16\% (DISK SEQ W) to 197\% (DISKB LAT), GCP variability varies from 0.08\% (DISK SEQ R) to 52\% (CPU DUR), and EGI variability ranges from 0.31\% (CPU LAT) to 155\% (DISK SEEK).
If we set a threshold on variability at 10\%, we can observe that AWS and GCP are the best providers with most (80\% and 67\%, respectively) of the benchmarks executed with a variability lower than the threshold. The result is worse with Azure (55\%), and EGI (54\%).
The results show that single metrics can only give a partial view of the problem.
For example, EGI obtains the best results for CPU- and network-related metrics (except for CPU TH LAT and NETB 1) and the worst ones in all the tests related to memory, in four out of eleven of the metrics used to measure disk performance and NETB 1.

While the standard deviation measures how dispersed the values are with respect to the mean, it does not addresses how data are distributed.\footnote{For example, values can be uniformly distributed between the minimum and maximum or most of the values can sit around the mean with few values far distant from it (also called outliers).}
To this end, and to measure the effect of the tails, we filtered the datasets with a pass-band filter, removing the first and last 5\% of the results, thus keeping the values between the $5^{th}$ and $95^{th}$ percentiles.
The results, not shown here for lack of room, show that the RSDs of the filtered datasets are on average 30\% smaller than the RSDs of the complete datasets: 35\% smaller for AWS, 26\% smaller for Azure and EGI, and 33\% smaller for GCP.
The difference among classes is negligible for all the providers but Azure, where the A* machines show a small difference (17\%), while the B* VMs a greater one (35\%). On the other hand, the mean values did not change.

Our analysis shows that if we consider a single provider, different resources are affected by different levels of variability.
The patterns of such variability appear to be difficult to understand through visual inspection and descriptive statistics.
To better understand the performance of single resources varies over time we conducted further experiments using ML methods, as described in Section~\ref{sec-ml}.

\subsection{Isolation}

\begin{figure}[t]
	\centering
	\includegraphics[width=0.75\linewidth]{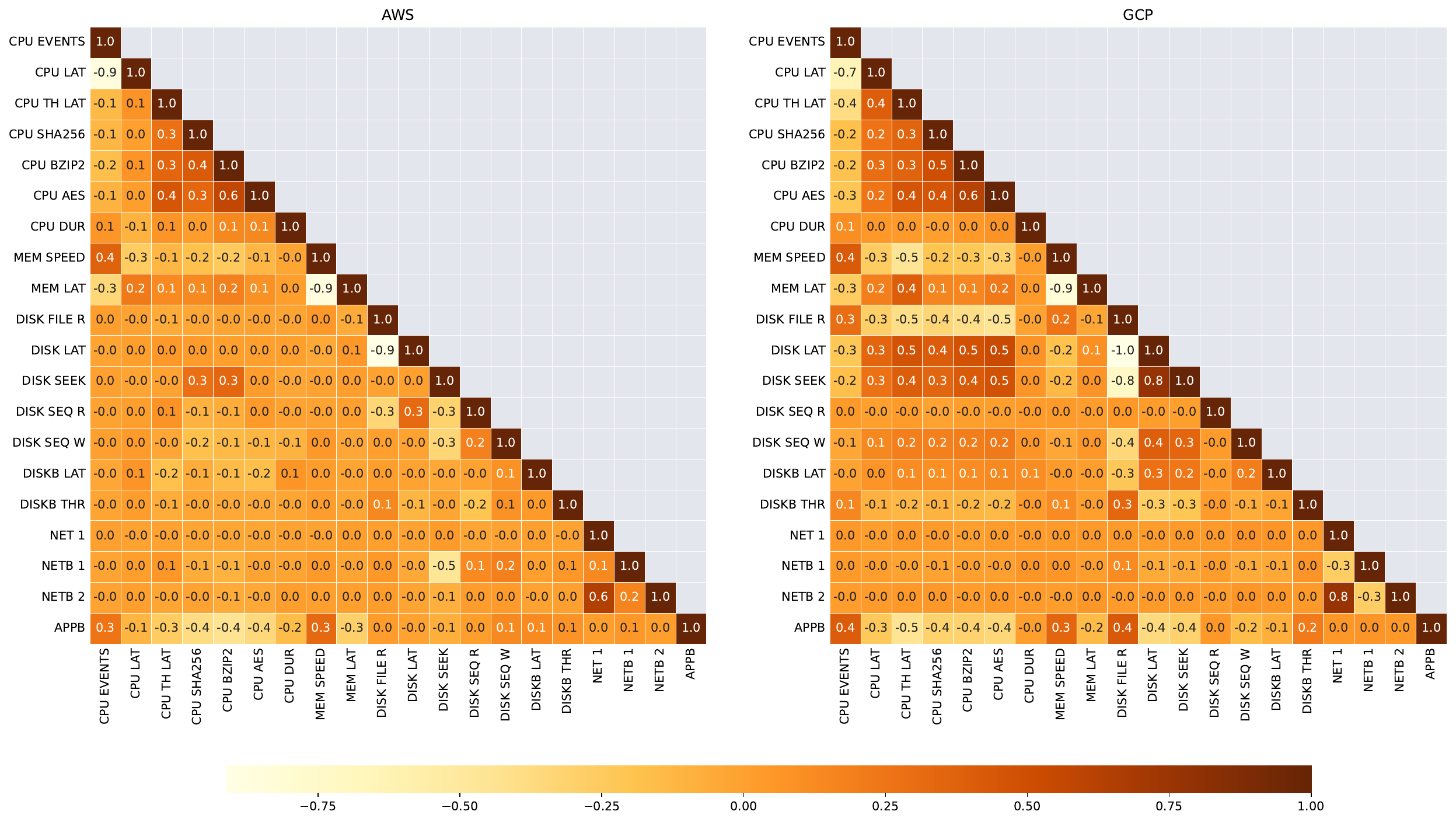}
	\caption{\nw{Correlation between resource metrics for AWS and GCP.}}
	\label{fig:correlation}
\end{figure}

\begin{figure}[t]
	\centering
	\includegraphics[width=0.6\linewidth]{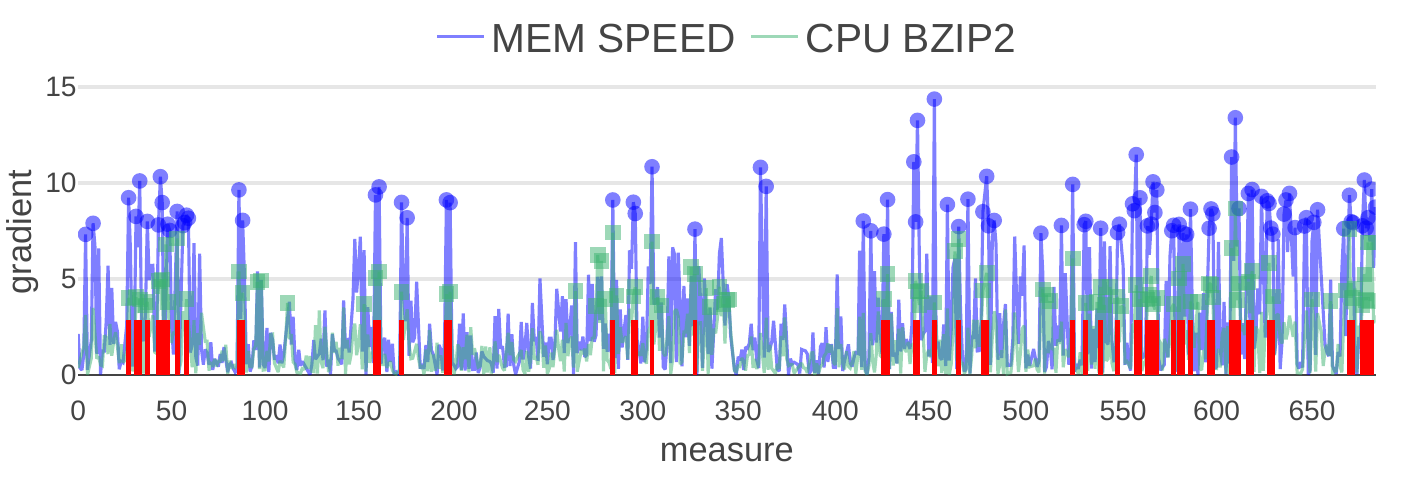}
	\caption{Comparison between CPU BZIP2 and MEM SPEED on Azure A2 VMs.}
	\label{fig:relations_azure_cpu_mem}
\end{figure}

While the resources of a VM work cooperatively at application level (e.g., if provisioned memory is not enough, the system starts swapping and application's performance degrades), to answer \rqiso we assessed whether the performance variability of a resource affects the others in the short term independently of the running application.
To do so, we employed two methods, the first based on correlation matrices and the second on gradient analysis. 

\nw{Figure~\ref{fig:correlation} shows the correlation matrices for AWS and GCP computed over the whole dataset and averaged over all the VMs types for each provider.
The other two providers yielded comparable results and their matrices are not included herein for the sake of brevity.
The matrix of AWS shows that most of the metrics are not correlated with one another (0.0 correlation). We can observe that some correlation exists between metrics that measure the same resource (e.g., CPU EVENTS and CPU LAT show a correlation equals to 0.9) but different resources appear to be almost completely independent with one another. 
The matrix of GCP shows a slightly higher correlation among the metrics.
For example, DISK LAT and DISK SEEK appear to be correlated with some CPU-related metrics (e.g., $0.5$ correlation with CPU AES). Similarly, MEM SPEED and MEM LAT show a correlation CPU-TH LAT of $-0.5$ and $0.4$, respectively.}

As for the gradient-based method, we computed the mean $\mu_m$ of all values ---collected over a month--- for each metric $m$.
For each value $x_m$, we then computed $x_m / \mu_m$ and the gradient between each pair of subsequent values to estimate the variations.
We took the times at which each metric showed the $n$ highest values of the gradient, that is, the $n$ time points with the most noticeable variations. We then compared the time points of every possible pair of metrics.
If the number of equal time points in the two sets is greater than $n \times t$, where $t$ is a threshold, a relationship between the changes of the two metrics (resources) exists.
We took $n=100$ and a $t=.6$ ($60\%$ similarity). Note that, to avoid capturing the natural dependency among resources at application level, we executed the benchmarks one after the other, within a short time window, to avoid parallel executions.

This analysis revealed that there is no noticeable dependence between single resources for AWS, GCP, and EGI VMs. 
Azure A2 VMs suggested a (weak) relationship between CPU BZIP2 and MEM SPEED, CPU BZIP2 and MEM OPS. Figure~\ref{fig:relations_azure_cpu_mem} shows the relationship between CPU BZIP2 and MEM SPEED. 61\% of the gradients, highlighted in the figure, coincide (red lines), and thus there exists a relationship between the two resources. We obtained a similar result while comparing CPU AES and DISK FILE W/R/F, CPU AES and DISK THR W/R, CPU AES and DISK SEQ W/R, abd CPU AES and MEM LAT.
%Moreover, there existed a relationship between CPU EVENTS and APPB values for B2MS. For B4MS the relationships were between APPB and MEM SPEED, APPB and CPU EVENTS. \gio{Qui parliamo di app, non abbiamo deciso di tagliare?} We did not find any relationship for GCP and EGI VMs.
%In contrast, as foreseen, there exists a relationship between the values obtained with APPB (complete application) and CPU BZIP2 / AES / TH LAT, APPB and DISK SEQ W for a1.xlarge and m5.xlarge VMs since resources cooperate at application level. 
%Figure~\ref{fig:relations_aws_appb_diskseqw} shows the relationship between APPB and DISK SEQ W: they share some 60\% (red fragments) of the top-100 most noticeable variations (blue dots and green squares).
%
%\begin{figure}[htbp]
%	\centering
%	\includegraphics[width=\linewidth]{imgs/relations/aws_a1.xlarge-1_DISK_SEQ_W_eb556_aws_a1.xlarge-1_APPB_7d7b0.pdf}
%	\caption{Comparison between APPB and DISK SEQ W for AWS - a1.xlarge.}
%	\label{fig:relations_aws_appb_diskseqw}
%\end{figure}

%These results show that application benchmarks, like APPB, exploit multiple resources (in this case, CPU and disk), and thus are not the best option to study performance variability of VMs. The implicit performance aggregation of these benchmarks does not allow one to understand the variability of single resources, but at least the experiment witnessed the ``implicit'' dependence between an application and the main resources it used. \gio{forse questa parte può confondere e andrebbe tolta.}
Overall, the results indicate that there is generally a weak correlation between resources in cloud providers.
This suggests that cloud providers are capable of managing resources in isolation.
Furthermore, these findings emphasize the importance of employing a multi-metric approach, as the observed correlations vary across different metrics and only specific pairs of metrics exhibit a significant level of correlation.

\subsection{Time}

\begin{figure}[h]
	\centering
	\begin{subfigure}[b]{0.49\linewidth}
		\includegraphics[width=\linewidth]{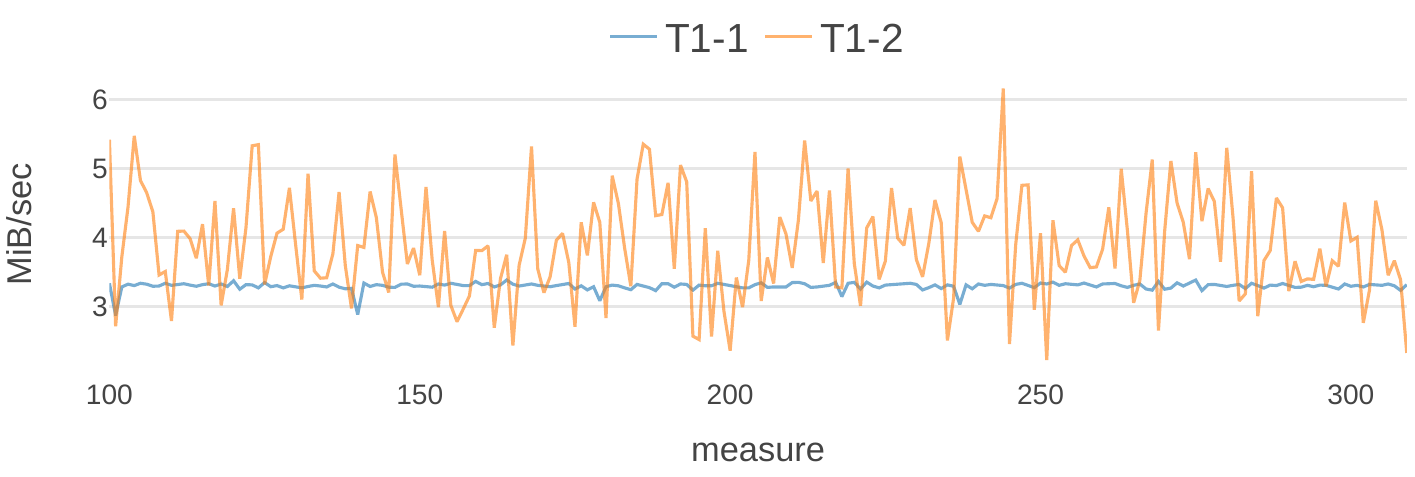}
		\caption{DISK THR R on EGI VMs.}
		\label{fig:egi_time_disk_thr_r}
	\end{subfigure}
	\hfill
	\begin{subfigure}[b]{0.49\linewidth}
		\includegraphics[width=\linewidth]{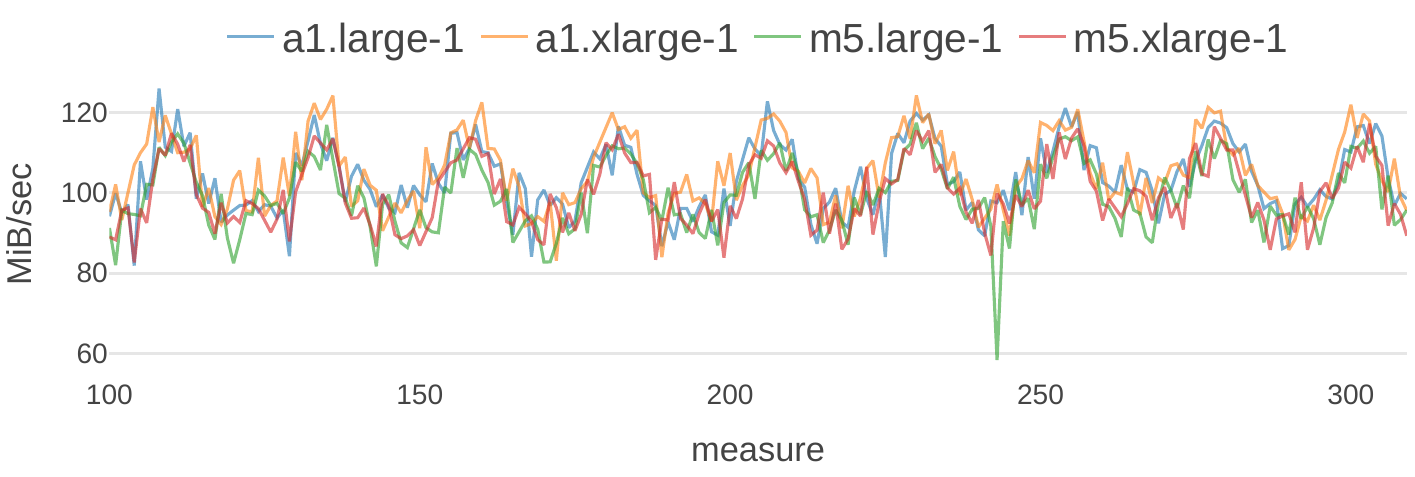}
		\caption{NET 1 on AWS VMs.}
		\label{fig:nench_netc}
	\end{subfigure}
	\caption{Time analysis (measure identifies the ordinal of retrieved values).}
	\label{fig:combined}
\end{figure}

Table~\ref{tab:bench_results_all_rsd_avg} witnesses that performance varies, with different breadths, no matter the resource one considers.
To answer \rqtime, we then decided to study the possible presence of variability patterns and cyclic behaviors. 
Our data analysis demonstrated that there are no apparent patterns between variations and particular hours, days, or weeks.
Figure~\ref{fig:egi_time_disk_thr_r} exemplifies a common case (DISKB THR R on EGI) where visual inspection does not allow to recognize any recurring pattern.
The performance appears to be highly fluctuating and unpredictable behavior ranging from a disk read speed of $2$ MiB/sec to more than $6$ MiB/sec.
However, we were able to observe a couple of highly cyclic behaviors. NET 1 and NETB 2 exhibited a recurring behavior: Figure~\ref{fig:nench_netc} shows NET 1 values for AWS VMs.
This should be related to the network performance of the download server since the other network benchmarks did not show a similar behavior.
The download speed is periodic and we got similar results also with Azure and GCP, while we were not able to appreciate any periodicity with EGI due to the limited download speed.

Our comprehensive analysis indicates that the majority of resources and providers do not exhibit noticeable recurring performance changes. However, we have identified a few exceptional cases where distinct patterns have emerged. In order to gain a deeper understanding of the temporal variations of individual resources, we conducted additional experiments using various ML methods, as elaborated in Section~\ref{sec-ml}.

\subsection{Cost}
\begin{figure}[t]
	\centering
	\begin{subfigure}[b]{0.49\linewidth}
		\includegraphics[width=\linewidth]{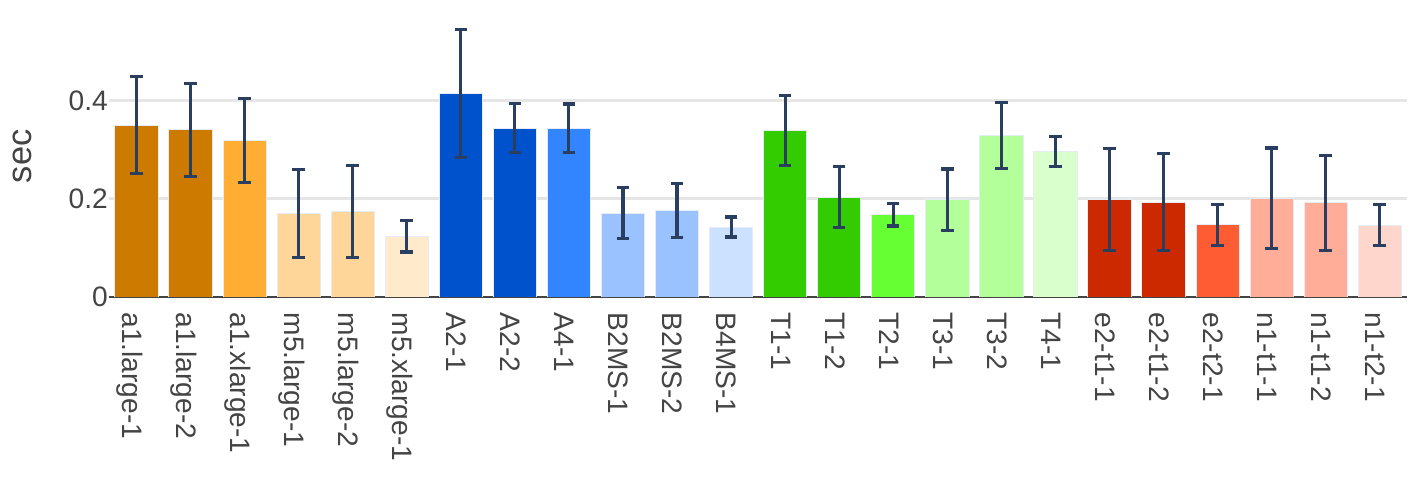}
		\caption{CPU DUR.}
		\label{fig:cpubench}
	\end{subfigure}
	\hspace{\fill}
	\begin{subfigure}[b]{0.49\linewidth}
		\includegraphics[width=\linewidth]{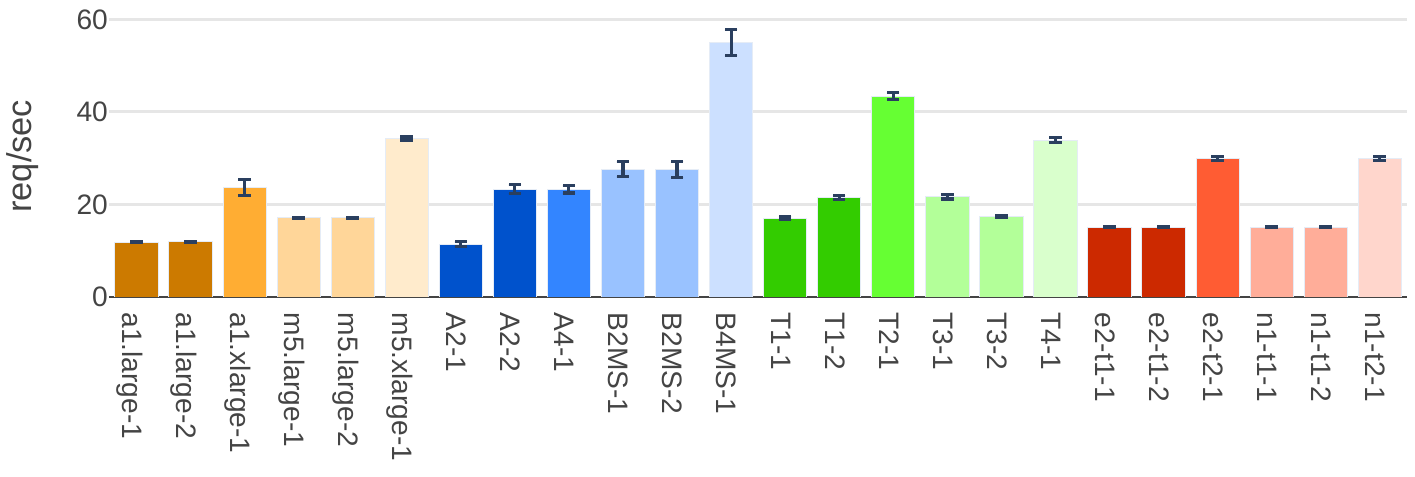}
		\caption{APPB.}
		\label{fig:webbench_rps}
	\end{subfigure}
	\caption{Performance among different VM types.}
	\label{fig:combined2}
\end{figure}

\rqcost started from the idea of studying whether more expensive VMs come with higher values, and thus whether investing more money always means getting better performance and stability. On average, this appears to be not true. 
Our experiments suggested that the performance of different VM types mainly depend on analyzed metrics and not on the type itself, nor the size. 
In this section, we provide some examples by presenting charts of the performance of different providers.
Besides the mean measured value, the black line on top of each bar shows the standard deviation.
%\begin{figure}[thbp]
%	\centering
%	\includegraphics[width=\linewidth]{imgs/bar/all_CPU_EVENTS_cd605.pdf}
%	\caption{CPU EVENTS.}
%	\label{fig:sysbench_cpu_eps}
%end{figure}
Figure~\ref{fig:cpubench} (CPU DUR) shows that the performance of VMs of different types is clearly not equal, but there is a small difference between machines of the same family, for AWS and Azure (e.g., a1 and m5, and A* and B*MS) and differences between different instances of the VMs.\footnote{Note that when we write \textit{VMtype/size-X}, \textit{X} is $1$ when we only have one instance, and can be $1$ or $2$ with two instances.}
GCP types (E2 and N1) are comparable, while the performance of EGI VMs was quite variable and there was no strong difference between VM types. 
Similarly, CPU EVENTS witnessed that the performance of a VM is not always proportional to the cost of the machine, and there is not always a sharp difference between families. For example, AWS a1 VMs obtained higher values than the more expensive m5 ones and the results for GCP E2 and N1 VMs are comparable.
EGI T1 and T3, and T2 and T4, provided similar results. On the other hand, there is a clear difference between different machine sizes, e.g., between AWS large and xlarge and GCP E2-T1 and E2-T2 types. 

%\begin{figure}[thbp]
%	\centering
%	\includegraphics[width=\linewidth]{imgs/bar/all_MEM_SPEED_11d45.pdf}
%	\caption{MEM SPEED - SysBench - Memory Speed [MiB/s] - (HIB)}
%	\label{fig:sysbench_memory_speed}
%\end{figure}

Figure \ref{fig:webbench_rps} refers to APPB and one can observe that the performance increased proportionally to the VM size for AWS a1 and m5 and for Azure B*MS VMs, but not for Azure A* machines where the results for A2 and A4 VMs were comparable.
In addition, AWS a1.xlarge VMs performed better than m5.large ones and they have the same cost per hour. The behaviors of GCP E2 and N1 VMs were comparable, while results with EGI VMs were variable and the highest performance was not of the VM with more resources.
Disk performance did not show any substantial difference between VMs of different sizes for AWS and GCP while there was a clear difference for Azure, and this is true for all Sysbench disk benchmarks. %Figure~\ref{fig:nench_dd_asws} presents the results we obtained with Nench disk benchmark (DISK SEQ W), which measure the average sequential write speed, and does not highlight any significant difference between VM types, with the exception of EGI, whose results are variable. 
DDBench (large block size) (DISKB THR) highlighted a noticeable difference only between AWS a1 and m5 VMs.
%(as shown in %Figure~\ref{fig:ddbench_large}).
%
% \begin{figure}[thbp]
% 	\centering
% 	\includegraphics[width=0.65\linewidth]{imgs/bar/all_DISK_LAT_6b10b.pdf}
% 	\caption{DISK LAT.}
% 	\label{fig:sysbench_fal}
% \end{figure}
%
%\begin{figure}[thbp]
%	\centering
%	\includegraphics[width=\linewidth]{imgs/bar/all_DISK_SEQ_W_eb556.pdf}
%	\caption{DISK SEQ W - Nench - DD, avg seq. write speed [MiB/s] (HIB)}
%	\label{fig:nench_dd_asws}
%\end{figure}
%
%\begin{figure}[thbp]
%	\centering
%	\includegraphics[width=\linewidth]{imgs/bar/all_DISKB_THR_61573.pdf}
%	\caption{DDBench - Large block size [MB/s] (HIB)}
%	\label{fig:ddbench_large}
%\end{figure}
Finally, the results for network benchmarks showed that download speed is independent of VM type and its cost.

\nw{To better relate performance and cost, we propose the cost/performance ratio (CPR) for five relevant metrics, one per resource type.
Table~\ref{tab:cost} shows the results (lower values are better). Being free of charge, we did not report any result for EGI.
GCP has the best value for CPU BZIP2 for classes C1 and C2, while it is less competitive with classes C3 and C4, where Azure and AWS are the best ones, respectively.
Looking at MEM SPEED, AWS has the best CPR with classes C1 and C2; its values are much higher than those of the competitors. Azure's memory performance is cheaper than AWS and EGI for C3 and C4 VMs.
The best CPR for DISK THR R is obtained by AWS that is above the competitors, while the CPRs related to networking are comparable, even if AWS is the best for small VMs and Azure for the bigger ones. 
Finally, if one considers APPB, GCP offered the best CPRs ratio for C1 and C2 VMs, and Azure for the bigger machines.
The CPR tells us that AWS is possibly the most cost-effective provider for small VMs, while one should prefer Azure for the more powerful ones.}

\begin{table*}[t]
	\renewcommand\arraystretch{1.3}
	\setlength{\tabcolsep}{7pt}
	\centering
 \footnotesize
	\begin{tabular}{r|r|c|c|c|c|c}
    &  & \textbf{CPU BZIP2} & \textbf{MEM SPEED} & \textbf{DISK THR R} & \textbf{NET 1} & \textbf{APPB} \\ \hline
		\multirow{3}{*}{\textbf{C1}} & \multicolumn{1}{c|}{AWS} & 558.72m & \cellcolor[HTML]{C0C0C0}2.78$\mu$ & \cellcolor[HTML]{C0C0C0}1.46m & \cellcolor[HTML]{C0C0C0}567.89$\mu$ & 4.89m \\
		& \multicolumn{1}{c|}{Azure}  & 1.19  & 6.32$\mu$ & 25.96m  & 615.82$\mu$ & 5.66m \\
		& \multicolumn{1}{c|}{GCP} &  \cellcolor[HTML]{C0C0C0}511.83m & 5.32$\mu$ & 2.91m  & 749.11$\mu$ &  4.71m 
		 \\ \hline
		\multirow{3}{*}{\textbf{C2}} & \multicolumn{1}{c|}{AWS} & 1.12 & 3.18$\mu$ & 2.70m  & 1.12m & 4.90m \\
		& \multicolumn{1}{c|}{Azure} &  \cellcolor[HTML]{EFEFEF}2.50 & 9.23$\mu$ & \cellcolor[HTML]{EFEFEF}54.06m  & 1.29m  & 7.85m \\
		& \multicolumn{1}{c|}{GCP} & 873.06m & 5.62$\mu$ & 5.70m &  1.50m & 4.76m \\
		 \hline
		\multirow{3}{*}{\textbf{C3}} & \multicolumn{1}{c|}{AWS} &  589.72m & 7.37$\mu$  & 2.72m & 1.16m  & 6.72m \\
		& \multicolumn{1}{c|}{Azure} &  574.05m  & 4.08$\mu$ & 8.00m & 904.81$\mu$  & \cellcolor[HTML]{C0C0C0}3.48m \\
		& \multicolumn{1}{c|}{GCP} &  876.03m & 9.30$\mu$ &  4.63m &  1.31m & 8.26m  \\ \hline
		\multirow{3}{*}{\textbf{C4}} & \multicolumn{1}{c|}{AWS} &  1.13  & 7.80$\mu$ & 5.65m & 2.32m & 6.71m \\
		& \multicolumn{1}{c|}{Azure} & 1.14  & 5.02$\mu$ &  15.97m & 1.87m & 3.49m \\
		& \multicolumn{1}{c|}{GCP} & 1.52  & \cellcolor[HTML]{EFEFEF}9.86$\mu$ & 9.57m &  \cellcolor[HTML]{EFEFEF}2.62m & \cellcolor[HTML]{EFEFEF}8.34m \\ \bottomrule
\end{tabular}
	\caption{Cost/performance ratio for a subset of considered metrics. Note that $m$ means $10^{-3}$, while $\mu$ means $10^{-6}$. The best VM for each metric is highlighted in dark gray, the worst in light gray.}
	\label{tab:cost}
\end{table*}

\section{Performance Predictive Models}
\label{sec-ml}

\nw{The data analysis presented in the previous section did not identify significant patterns within the VM performance data collected.
In particular, the preliminary answers to \rqres (relative to the stability of the performance of common VMs) and \rqtime (on the relationship between offered and time of the day or day of the week) suggest that VM performances are unstable and very difficult to predict regardless of the provider.
%Despite the variability of VMs being confirmed by multiple works from literature, none of them tried to study temporal VM performance data using advanced ML techniques.
In this section, we propose an in-depth analysis of data collected using multiple ML models for forecasting and classification.
Specifically, we describe the following two experiments: time-series forecasting to predict VMs performance based on past observations, and a series of multivariate prediction tasks based on classification to detect recurring temporal trends.\footnote{The code for replicating the experiments can be downloaded from: \url{https://doi.org/10.5281/zenodo.8014668}.}}

\nw{The main objective of the first experiment is to consider and study individually the predictability of the performance of each measured resource for each of the four providers (AWS, Azure, GCP, EGI).
This analysis allowed us to give a better answer to \rqres, considering that resources correctly predicted by a ML model can be considered relatively stable, or at least regular and cyclic from a temporal point of view.
Given our data analysis and previous results from literature, we are aware of the high degree of variability of VM performances, but thanks to the quantity and diversity of the collected data, we believe that there is enough material to extract new knowledge and adequately train multiple ML models for resource performance forecasting.}
%From this experiment, we expect to gain important insights and to highlight differences in the forecasting accuracy depending on the VM and provider under analysis.

\nw{The goal of the second experiment is to understand whether a classification model is able to associate a given set of performance values to a specific time period, either within the day (time of the day), or within the week (day of the week).
The results of this experiment allowed us to give a more precise answer to \rqtime.
Accurate predictions from the models would highlight a relationship between the performance of a VM and different times of day (or days of the week), while poor performance would further confirm the unpredictability of VM performance.}

\subsection{Time-Series Forecasting}
\nw{The time-series forecasting experiment required a first step of pre-processing to \textit{a)} normalize data, \textit{b)} fill missing values and \textit{c)} make data stationarity through a differencing process that stabilizes the mean and variance across time. %make data stationary through a  differencing process, that is, making the mean and variance stable across time.
%\footnote{Differencing consists in computing the difference between consecutive observations to remove seasonal variations (seasonality) and make the time series stationary, i.e., with a stable mean and variance across time.}.
We normalized data with a min-max approach, filled missing values with a simple mean-based imputation technique, and finally performed a first-order differencing to obtain a stationary series.
As confirmed by the augmented Dickey–Fuller test \cite{dickey1979distribution}, a further second or higher order of differencing was not needed for our data.}

\nw{For this experiment, we considered all the four classes of VM (C1, C2, C3, C4) for each of the four providers (AWS, Azure, GCP, EGI), for a total of 16 tests, each involving all the 28 resource metrics collected by \benchSuite.
These metrics were considered separately, since forecasting tasks look at one attribute at a time.
We fitted and tested three different models, namely Vector AutoRegression (VAR) \cite{lutkepohl2005new}, Autoregressive Integrated Moving Average (ARIMA) \cite{box2015time}, and Seasonal AutoRegressive Integrated Moving Average with eXogenous regressors model (SARIMAX) \cite{durbin2012time}.
ARIMA and SARIMAX are more advanced methods from the family of moving average models, and SARIMAX in particular is intended to better manage seasonality. For this reason, we expect them to obtain better results compared to VAR.
The performance data used for fitting the models were observed at intervals of one hour each, for a total of consecutive observations for each resource around 700--800, depending on the VM provider.
We set the number of subsequent data points to predict to 5.}

\nw{To evaluate the resulting models we considered two regression loss metrics commonly used to measure the accuracy of forecasts: Mean Absolute Error (MAE) %, mean squared error (MSE), mean absolute percentage error (MAPE)
and Mean Absolute Scaled Error (MASE) \cite{hyndman2006another}.
MAE computes the average absolute error between the real and the predicted values:
\begin{displaymath}
    \frac{1}{N}\sum_{i=1}^{N}|x_i^{real}-x_i^{pred}| \;
\end{displaymath}
\noindent The closer the result is to 0, the higher is the accuracy of the forecasts. 
It is a popular and immediate metric, whose main limitation is being hard to compare between different datasets and attributes.
To overcome this problem, we leveraged the more advanced MASE, a scaled indicator that compares the error of a model's predictions with those from a naive model:
\begin{displaymath}
    \frac{1}{N}\sum_{i=1}^{N}\frac{|x_i^{real}-x_i^{pred}|}{MAE(x^{real},x^{naive})} \;
\end{displaymath}
\noindent Again, the closer the result is to 0, the higher is the accuracy; in particular, an error <1 means that the predictions are more accurate compared to the naive model.
This allowed us to compare the results across VMs and between the different resource metrics collected by \benchSuite.}

\nw{Table \ref{tab:forecasting}  reports the detailed MASE of the predictions obtained with the ARIMA model, i.e., the model that obtained the best results among the three.
The values in the last column and the last row show, respectively, the average MASE by performance metric, and the average MASE by VM. The first allows to identify the more predictable performance metric, regardless of the provider: memory appears to be the more predictable resource, although it comprises only two benchmark metrics, and especially so thanks to the extremely low MASE of AWS and GCP machines also looking at MAE; CPU is also quite predictable, and this is confirmed by the very low MAE in a good number of cases, particularly with AWS.
On the other hand, Disk and Net metrics obtained contrasting and generally worse results, with many occurrences of MASE >1.
From the point of view of providers, AWS appears predictable in many cases, e.g., with Memory (MEM SPEED, MEM LAT), but not much with Network, especially with NET 1, NET 3, NET 4 and NET 5.
EGI also appears hard to predict with Disk metrics, but obtains good results regarding CPU (CPU LAT, CPU TH LAT, CPU DUR).
On the other hand, GCP appears more easily predictable on Disk and Memory metrics, particularly with VMs of class C2.
In general, the more predictable is GCP with an average 0.62 MASE, followed by AWS and EGI with 0.74 and 0.77, while Azure appears to be way less predictable (1.14).
On the other hand, there is no clear difference between classes of VMs, with smaller sized machines being a little more predictable (C1 has an average MASE of 0.80 and C2 of 0.66) compared to bigger sized VMs (0.84 for C3 and 0.97 for C4).}
%74-114-61.5-76.75
%79.5-65.5-84-97.25

\nw{Moreover, Table \ref{tab:forecasting_overall} reports the average MASE and MAE computed over all the considered resource metrics for each VM, for all the three forecasting models tested.
Results show that SARIMAX is on par with ARIMA, while the more simple VAR model exhibits a larger error.
The general trends of MASE for providers and VM type are invariant with respect to the employed model. 
Additionally, MAE is stable regardless of the model, although this evaluation metric is less precise compared to MASE.}

\begin{table*}[t]
\renewcommand\arraystretch{1.3}
\setlength{\tabcolsep}{3.4pt}
\fontsize{8}{9.6}\selectfont
\centering
\begin{tabular}{@{}r|cccc|cccc|cccc|cccc|c@{}}
& \multicolumn{4}{c|}{\textbf{AWS}} & \multicolumn{4}{c|}{\textbf{Azure}} & \multicolumn{4}{c|}{\textbf{GCP}} & \multicolumn{4}{c|}{\textbf{EGI}} & \\ \hline
\textbf{Metric} & C1 & C2 & C3 & C4 & C1 & C2 & C3 & C4 & C1 & C2 & C3 & C4 & C1 & C2 & C3 & C4 & \textbf{Overall} \\ \hline
CPU EVENTS & 0.50  \phantom{-} &0.67 \phantom{-} &0.46 \textsuperscript{\textdagger} & 0.44 \textsuperscript{\textdagger} & \cellcolor[HTML]{EFEFEF}1.48 \phantom{-} &1.22 \phantom{-} &0.41 \phantom{-} &0.70  \phantom{-} &0.8  \phantom{-} &0.23 \textsuperscript{\textdagger} & 0.55 \textsuperscript{\textdagger} & \cellcolor[HTML]{C0C0C0}0.14 \textsuperscript{\textdaggerdbl} & 0.37 \textsuperscript{\textdagger} & 0.42 \phantom{-} &0.46 \phantom{-} &1.39 \phantom{-} &0.64 \\
CPU LAT & 0.54 \phantom{-} &0.34 \phantom{-} &0.67 \textsuperscript{\textdagger} & 0.39 \phantom{-} &\cellcolor[HTML]{EFEFEF}1.24 \phantom{-} &1.16 \phantom{-} &0.11 \textsuperscript{\textdaggerdbl} & 1.06 \phantom{-} &0.43 \phantom{-} &0.44 \textsuperscript{\textdaggerdbl} & 0.62 \phantom{-} &\cellcolor[HTML]{C0C0C0}0.00  \textsuperscript{\textdaggerdbl} & 1.06 \phantom{-} &\cellcolor[HTML]{C0C0C0}0.00  \textsuperscript{\textdaggerdbl} & 0.51 \phantom{-} &4.08 \phantom{-} &0.79 \\
CPU TH LAT & 0.95 \phantom{-} &0.37 \textsuperscript{\textdaggerdbl} & 0.56 \phantom{-} &0.63 \phantom{-} &1.15 \phantom{-} &0.88 \phantom{-} &0.89 \phantom{-} &\cellcolor[HTML]{EFEFEF}1.30  \phantom{-} &1.06 \phantom{-} &0.51 \textsuperscript{\textdagger} & 0.71 \phantom{-} &0.44 \phantom{-} &\cellcolor[HTML]{C0C0C0}0.31 \textsuperscript{\textdagger} & 0.46 \phantom{-} &0.54 \phantom{-} &0.77 \phantom{-} &0.72 \\
CPU SHA256  & 0.71 \phantom{-} &\cellcolor[HTML]{C0C0C0}0.20  \textsuperscript{\textdaggerdbl} & 0.67 \phantom{-} &0.24 \textsuperscript{\textdaggerdbl} & 1.74 \phantom{-} &1.00  \phantom{-} &0.49 \phantom{-} &\cellcolor[HTML]{EFEFEF}1.75 \phantom{-} &0.83 \phantom{-} &0.53 \phantom{-} &0.52 \phantom{-} &0.77 \phantom{-} &0.42 \phantom{-} &0.88 \phantom{-} &0.51 \phantom{-} &0.59 \phantom{-} &0.74 \\
CPU BZIP2  & 0.48 \textsuperscript{\textdagger} & \cellcolor[HTML]{C0C0C0}0.14 \textsuperscript{\textdaggerdbl} & 0.43 \textsuperscript{\textdaggerdbl} & 0.66 \phantom{-} &1.62 \phantom{-} &0.78 \phantom{-} &0.34 \textsuperscript{\textdagger} & \cellcolor[HTML]{EFEFEF}2.51 \phantom{-} &0.65 \phantom{-} &0.43 \phantom{-} &0.68 \phantom{-} &1.03 \phantom{-} &1.22 \phantom{-} &0.41 \phantom{-} &1.24 \phantom{-} &1.31 \phantom{-} &0.87 \\
CPU AES  & \cellcolor[HTML]{EFEFEF}2.22 \phantom{-} &\cellcolor[HTML]{C0C0C0}0.12 \textsuperscript{\textdaggerdbl} & 0.42 \textsuperscript{\textdagger} & 0.37 \textsuperscript{\textdaggerdbl} & 1.10  \phantom{-} &0.99 \phantom{-} &0.95 \phantom{-} &1.76 \phantom{-} &0.40  \textsuperscript{\textdagger} & 1.39 \phantom{-} &0.65 \phantom{-} &0.44 \textsuperscript{\textdagger} & 0.55 \phantom{-} &0.90  \phantom{-} &0.48 \phantom{-} &1.33 \phantom{-} &0.88 \\
CPU DUR & 0.73 \phantom{-} &0.60  \phantom{-} &0.52 \phantom{-} &0.75 \phantom{-} &0.49 \phantom{-} &0.49 \phantom{-} &0.68 \phantom{-} &\cellcolor[HTML]{EFEFEF}1.55 \phantom{-} &0.74 \phantom{-} &0.93 \phantom{-} &0.69 \phantom{-} &1.04 \phantom{-} &1.01 \phantom{-} &\cellcolor[HTML]{C0C0C0}0.41 \phantom{-} &0.92 \phantom{-} &0.57 \phantom{-} &0.76 \\ \hline
MEM SPEED & \cellcolor[HTML]{C0C0C0}0.10  \textsuperscript{\textdaggerdbl} & 0.37 \textsuperscript{\textdagger} & 0.09 \textsuperscript{\textdaggerdbl} & 0.68 \phantom{-} &\cellcolor[HTML]{EFEFEF}1.72 \phantom{-} &1.04 \phantom{-} &0.81 \phantom{-} &0.62 \phantom{-} &0.26 \textsuperscript{\textdaggerdbl} & 0.82 \phantom{-} &0.37 \textsuperscript{\textdaggerdbl} & 0.69 \phantom{-} &0.22 \phantom{-} &0.71 \phantom{-} &0.81 \phantom{-} &0.55 \phantom{-} &0.62 \\
MEM LAT & 0.05 \textsuperscript{\textdaggerdbl} & 0.45 \textsuperscript{\textdagger} & \cellcolor[HTML]{C0C0C0}0.00  \textsuperscript{\textdaggerdbl} & 0.31 \phantom{-} &\cellcolor[HTML]{EFEFEF}1.41 \phantom{-} &1.11 \phantom{-} &0.83 \phantom{-} &0.62 \phantom{-} &\cellcolor[HTML]{C0C0C0}0.00  \textsuperscript{\textdaggerdbl} & 1.05 \phantom{-} &\cellcolor[HTML]{C0C0C0}0.00  \textsuperscript{\textdaggerdbl} & 0.73 \phantom{-} &0.29 \phantom{-} &0.77 \phantom{-} &0.73 \phantom{-} &0.40  \phantom{-} &0.55 \\ \hline
DISK FILE R & 1.04 \phantom{-} &0.54 \phantom{-} &0.61 \phantom{-} &0.87 \phantom{-} &1.17 \phantom{-} &0.64 \phantom{-} &2.82 \phantom{-} &\cellcolor[HTML]{EFEFEF}3.75 \phantom{-} &0.39 \phantom{-} &\cellcolor[HTML]{C0C0C0}0.19 \textsuperscript{\textdagger} & 1.21 \phantom{-} &0.62 \phantom{-} &0.74 \phantom{-} &1.27 \phantom{-} &1.16 \phantom{-} &0.48 \phantom{-} &1.09 \\
DISK FILE W & 1.04 \phantom{-} &0.54 \phantom{-} &0.61 \phantom{-} &0.87 \phantom{-} &1.17 \phantom{-} &0.64 \phantom{-} &2.82 \phantom{-} &\cellcolor[HTML]{EFEFEF}3.75 \phantom{-} &0.39 \phantom{-} &\cellcolor[HTML]{C0C0C0}0.19 \textsuperscript{\textdagger} & 1.21 \phantom{-} &0.62 \phantom{-} &0.74 \phantom{-} &1.27 \phantom{-} &1.16 \phantom{-} &0.48 \phantom{-} &1.09 \\
DISK FILE F & 1.04 \phantom{-} &0.54 \phantom{-} &0.61 \phantom{-} &0.87 \phantom{-} &1.17 \phantom{-} &0.64 \phantom{-} &2.82 \phantom{-} &\cellcolor[HTML]{EFEFEF}3.74 \phantom{-} &0.39 \phantom{-} &\cellcolor[HTML]{C0C0C0}0.19 \textsuperscript{\textdagger} & 1.21 \phantom{-} &0.62 \phantom{-} &0.69 \phantom{-} &1.27 \phantom{-} &1.16 \phantom{-} &0.48 \phantom{-} &1.09 \\
DISK THR R & 1.04 \phantom{-} &0.54 \phantom{-} &0.61 \phantom{-} &0.87 \phantom{-} &1.17 \phantom{-} &0.64 \phantom{-} &2.83 \phantom{-} &\cellcolor[HTML]{EFEFEF}3.59 \phantom{-} &0.39 \phantom{-} &\cellcolor[HTML]{C0C0C0}0.19 \textsuperscript{\textdagger}   & 1.21 \phantom{-} &0.62 \phantom{-} &0.74 \phantom{-} &1.27 \phantom{-} &1.16 \phantom{-} &0.48 \phantom{-} &1.08 \\
DISK THR W & 1.08 \phantom{-} &0.54 \phantom{-} &0.61 \phantom{-} &0.87 \phantom{-} &1.17 \phantom{-} &0.64 \phantom{-} &2.83 \phantom{-} &\cellcolor[HTML]{EFEFEF}3.75 \phantom{-} &0.39 \phantom{-} &\cellcolor[HTML]{C0C0C0}0.19 \textsuperscript{\textdagger}   & 1.21 \phantom{-} &0.62 \phantom{-} &0.74 \phantom{-} &1.28 \phantom{-} &1.16 \phantom{-} &0.48 \phantom{-} &1.10 \\
DISK LAT & 1.02 \phantom{-} &\cellcolor[HTML]{C0C0C0}0.00  \textsuperscript{\textdaggerdbl}   &0.99 \phantom{-} &0.54 \phantom{-} &1.17 \phantom{-} &0.67 \phantom{-} &2.90  \phantom{-} &\cellcolor[HTML]{EFEFEF}4.27 \phantom{-} &0.40  \phantom{-} &0.25 \phantom{-} &1.39 \phantom{-} &0.67 \phantom{-} &0.54 \phantom{-} &1.37 \phantom{-} &0.79 \phantom{-} &0.60  \phantom{-} &1.10 \\
DISK SEEK & -- \phantom{-} &-- \phantom{-} &0.93 \textsuperscript{\textdaggerdbl}   &\cellcolor[HTML]{C0C0C0}0.14 \textsuperscript{\textdaggerdbl}   &0.87 \phantom{-} &0.24 \textsuperscript{\textdagger}   & 0.18 \textsuperscript{\textdaggerdbl}   &0.61 \phantom{-} &0.30  \textsuperscript{\textdagger}   & 0.21 \textsuperscript{\textdagger}   & 1.01 \phantom{-} &1.12 \phantom{-} &1.09 \phantom{-} &\cellcolor[HTML]{EFEFEF}1.74 \phantom{-} &0.71 \phantom{-} &1.47 \phantom{-} &0.66 \\
DISK SEQ R & -- \phantom{-} &-- \phantom{-} &1.07 \phantom{-} &\cellcolor[HTML]{C0C0C0}0.26 \textsuperscript{\textdaggerdbl}   &0.95 \phantom{-} &0.52 \phantom{-} &\cellcolor[HTML]{EFEFEF}1.64 \phantom{-} &0.87 \phantom{-} &0.40  \textsuperscript{\textdaggerdbl}   &0.60  \textsuperscript{\textdaggerdbl}   &0.32 \textsuperscript{\textdaggerdbl}   &0.31 \textsuperscript{\textdaggerdbl}   &1.04 \phantom{-} &1.60  \phantom{-} &0.74 \phantom{-} &1.22 \phantom{-} &0.72 \\
DISK SEQ W  & 1.11 \phantom{-} &0.11 \textsuperscript{\textdaggerdbl}   &0.23 \textsuperscript{\textdaggerdbl}   &\cellcolor[HTML]{C0C0C0}0.02 \textsuperscript{\textdaggerdbl}   &1.07 \phantom{-} &0.25 \textsuperscript{\textdagger}   & 0.61 \phantom{-} &0.72 \textsuperscript{\textdagger}   & 0.77 \phantom{-} &0.36 \phantom{-} &0.60  \phantom{-} &0.40  \phantom{-} &1.02 \phantom{-} &0.82 \phantom{-} &\cellcolor[HTML]{EFEFEF}1.13 \phantom{-} &0.72 \phantom{-} &0.62 \\
DISKB LAT & 0.89 \phantom{-} &0.65 \phantom{-} &0.29 \phantom{-} &0.68 \phantom{-} &0.91 \phantom{-} &1.25 \phantom{-} &\cellcolor[HTML]{EFEFEF}1.38 \phantom{-} &1.32 \phantom{-} &0.45 \phantom{-} &\cellcolor[HTML]{C0C0C0}0.13 \textsuperscript{\textdagger}   & 0.36 \phantom{-} &0.27 \phantom{-} &1.16 \phantom{-} &0.29 \phantom{-} &1.05 \phantom{-} &1.06 \phantom{-} &0.76 \\
DISKB THR & 0.65 \phantom{-} &0.23 \phantom{-} &\cellcolor[HTML]{EFEFEF}1.19 \phantom{-} &0.62 \phantom{-} &0.18 \textsuperscript{\textdaggerdbl}   &\cellcolor[HTML]{C0C0C0}0.16 \textsuperscript{\textdaggerdbl}   &0.46 \phantom{-} &1.06 \phantom{-} &0.43 \phantom{-} &0.50  \phantom{-} &0.27 \phantom{-} &0.52 \phantom{-} &0.64 \phantom{-} &0.25 \textsuperscript{\textdaggerdbl}   &0.42 \textsuperscript{\textdagger}   & 0.29 \textsuperscript{\textdaggerdbl}   &0.49 \\ \hline
NET 1  & 0.77 \phantom{-} &\cellcolor[HTML]{EFEFEF}2.38 \phantom{-} &0.81 \phantom{-} &1.02 \phantom{-} &0.72 \phantom{-} &0.81 \phantom{-} &0.74 \phantom{-} &1.04 \phantom{-} &0.50  \phantom{-} &0.53 \phantom{-} &0.61 \phantom{-} &\cellcolor[HTML]{C0C0C0}0.43 \textsuperscript{\textdagger}   & 0.63 \phantom{-} &\cellcolor[HTML]{C0C0C0}0.43 \phantom{-} &0.50  \phantom{-} &0.39 \phantom{-} &0.77 \\
NET 2  & 0.89 \phantom{-} &\cellcolor[HTML]{C0C0C0}0.20  \phantom{-} &0.51 \phantom{-} &0.64 \phantom{-} &0.60  \phantom{-} &0.44 \phantom{-} &0.95 \phantom{-} &0.77 \phantom{-} &0.76 \phantom{-} &1.04 \phantom{-} &0.57 \phantom{-} &0.80  \phantom{-} &0.50  \phantom{-} &0.65 \phantom{-} &0.63 \phantom{-} &\cellcolor[HTML]{EFEFEF}1.10  \phantom{-} &0.69 \\
NET 3  & 1.25 \phantom{-} &1.47 \phantom{-} &\cellcolor[HTML]{EFEFEF}1.70  \phantom{-} &1.20  \phantom{-} &0.67 \phantom{-} &0.60  \phantom{-} &0.53 \phantom{-} &0.54 \phantom{-} &0.48 \phantom{-} &\cellcolor[HTML]{C0C0C0}0.22 \textsuperscript{\textdagger}   & 0.38 \phantom{-} &0.76 \phantom{-} &0.63 \textsuperscript{\textdagger}   & 0.52 \phantom{-} &0.50  \phantom{-} &0.35 \phantom{-} &0.74 \\
NET 4  & 0.63 \phantom{-} &2.05 \phantom{-} &1.79 \phantom{-} &\cellcolor[HTML]{EFEFEF}2.14 \phantom{-} &0.74 \phantom{-} &0.72 \phantom{-} &0.52 \phantom{-} &0.90  \phantom{-} &1.28 \phantom{-} &\cellcolor[HTML]{C0C0C0}0.13 \textsuperscript{\textdaggerdbl}   &1.15 \phantom{-} &1.05 \phantom{-} &0.91 \phantom{-} &0.35 \textsuperscript{\textdagger}   & 0.75 \phantom{-} &0.50  \phantom{-} &0.97 \\
NET 5  & 2.01 \phantom{-} &3.09 \phantom{-} &2.38 \phantom{-} &\cellcolor[HTML]{EFEFEF}4.17 \phantom{-} &0.48 \textsuperscript{\textdaggerdbl}   &0.38 \textsuperscript{\textdaggerdbl}   &\cellcolor[HTML]{C0C0C0}0.27 \textsuperscript{\textdaggerdbl}   &0.39 \textsuperscript{\textdagger}   & 0.82 \phantom{-} &1.21 \phantom{-} &0.32 \phantom{-} &0.59 \phantom{-} &1.21 \phantom{-} &0.60  \textsuperscript{\textdagger}   & 0.61 \textsuperscript{\textdagger}   & 0.98 \textsuperscript{\textdagger}   & 1.22 \\
NETB 1 & 0.62 \textsuperscript{\textdagger}   & 1.17 \phantom{-} &\cellcolor[HTML]{C0C0C0}0.36 \textsuperscript{\textdagger}   & 0.41 \textsuperscript{\textdagger}   & \cellcolor[HTML]{EFEFEF}1.74 \phantom{-} &0.72 \phantom{-} &0.75 \phantom{-} &0.37 \phantom{-} &1.31 \phantom{-} &0.97 \phantom{-} &0.69 \phantom{-} &1.71 \phantom{-} &-- \phantom{-} &-- \phantom{-} &-- \phantom{-} &-- \phantom{-} &0.68 \\
NETB 2 & 0.63 \phantom{-} &1.23 \phantom{-} &0.56 \phantom{-} &0.54 \phantom{-} &0.97 \phantom{-} &0.41 \phantom{-} &1.04 \phantom{-} &0.88 \phantom{-} &0.72 \phantom{-} &\cellcolor[HTML]{C0C0C0}0.37 \phantom{-} &1.08 \phantom{-} &0.60  \phantom{-} &0.55 \phantom{-} &0.47 \textsuperscript{\textdagger}   & 0.44 \phantom{-} &\cellcolor[HTML]{EFEFEF}1.40  \phantom{-} &0.74 \\ \hline
APPB & 0.96 \phantom{-} &0.34 \textsuperscript{\textdagger}   & \cellcolor[HTML]{C0C0C0}0.06 \textsuperscript{\textdaggerdbl}   &0.13 \textsuperscript{\textdaggerdbl}   &\cellcolor[HTML]{EFEFEF}1.72 \phantom{-} &0.63 \phantom{-} &0.56 \phantom{-} &1.15 \phantom{-} &0.16 \textsuperscript{\textdagger}   & 0.36 \textsuperscript{\textdagger}   & 1.07 \textsuperscript{\textdagger}   & 0.43 \textsuperscript{\textdagger}   & 0.46 \phantom{-} &0.30  \phantom{-} &0.91 \phantom{-} &0.84 & 0.63 \\
\hline
\textbf{Overall}    & 0.82 & 0.67 & 0.71 & 0.76 & 1.09 & 0.70  & 1.15 & 1.62 & 0.57 & 0.51 & 0.74 & 0.64 & 0.70  & 0.74 & 0.76 & 0.87 & \\ \bottomrule
\end{tabular}
\caption{\nw{Detailed results of time-series forecasting for the ARIMA model, showing for each resource metric and each VM the corresponding MASE of the prediction. \textsuperscript{\textdagger}: associated MAE < 0.05, \textsuperscript{\textdaggerdbl}: associated MAE < 0.02. The best result for each VM class of each provider is highlighted in dark gray, the worst in light gray.}}
\label{tab:forecasting}
\end{table*}

\begin{table*}[t]
\renewcommand\arraystretch{1.4}
\setlength{\tabcolsep}{4.2pt}
\fontsize{8}{9.6}\selectfont
\centering
\begin{tabular}{@{}r|r|cccc|cccc|cccc|cccc@{}}
\multicolumn{2}{c|}{} & \multicolumn{4}{c|}{\textbf{AWS}} & \multicolumn{4}{c|}{\textbf{Azure}} & \multicolumn{4}{c|}{\textbf{GCP}} & \multicolumn{4}{c}{\textbf{EGI}} \\ \hline
\textbf{Model} & \textbf{Metric} & C1 & C2 & C3 & C4 & C1 & C2 & C3 & C4 & C1 & C2 & C3 & C4 & C1 & C2 & C3 & C4 \\ \hline
\multirow{2}{*}{VAR} & MASE & 1.093 & 0.954 & 1.213 & 0.963 & 1.522 & 1.186 & 1.494 & \cellcolor[HTML]{EFEFEF}2.100  & 0.925 & \cellcolor[HTML]{C0C0C0}0.755 & 1.005 & 0.782 & 0.777 & 1.055 & 0.922 & 0.925 \\
& MAE & 0.104 & \cellcolor[HTML]{C0C0C0}0.089 & 0.108 & 0.097 & \cellcolor[HTML]{EFEFEF}0.258 & 0.207 & 0.142 & 0.196 & 0.118 & 0.093 & 0.133 & 0.096 & 0.100  & 0.165 & 0.144 & 0.120 \\ \hline
\multirow{2}{*}{ARIMA} & MASE & 0.819 & 0.674 & 0.705 & 0.761 & 1.091 & 0.701 & 1.149 & \cellcolor[HTML]{EFEFEF}1.619 & 0.569 &\cellcolor[HTML]{C0C0C0} 0.506 & 0.738 & 0.644 & 0.696 & 0.740 & 0.756 & 0.868 \\
& MAE & 0.110 & \cellcolor[HTML]{C0C0C0}0.081 & 0.097 & 0.103 & \cellcolor[HTML]{EFEFEF}0.257 & 0.163 & 0.163 & 0.220 & 0.113 & 0.089 & 0.141 & 0.133 & 0.123 & 0.178 & 0.184 & 0.167 \\ \hline
\multirow{2}{*}{SARIMAX} & MASE & 0.810 & 0.761 & 0.764 & 0.783 & 1.224 & 0.607 & 1.226 & \cellcolor[HTML]{EFEFEF}1.624 & 0.527 & \cellcolor[HTML]{C0C0C0}0.473 & 0.782 & 0.643 & 0.701 & 0.726 & 0.798 & 0.854 \\
& MAE & 0.113 & 0.093 & 0.103 & 0.108 & \cellcolor[HTML]{EFEFEF}0.289 & 0.136 & 0.175 & 0.221 & 0.102 & \cellcolor[HTML]{C0C0C0}0.087 & 0.147 & 0.132 & 0.129 & 0.170 & 0.197 & 0.161 \\ \bottomrule
\end{tabular}
\caption{\nw{Overall time-series forecasting results for performance prediction with time-series forecasting models, showing both MASE and MAE for all the three algorithms tested. Best results in dark gray, worst in light gray.}}
\label{tab:forecasting_overall}
\end{table*}

\subsection{Multivariate Predictive Models}
\nw{To assess whether or not VM performance is dependent on the time of the day or the day of the week (\rqtime), we defined three multivariate predictive tasks, namely \textit{TimeDay}, \textit{DayWeek}, and \textit{Weekend}.
All these tasks were designed as classification problems and are trained on the entire dataset of observed resource metrics associated with a given VM provider, with each entry accompanied by a label representing a temporal attribute.
The goal of the trained model is to learn an association between the set of resource metrics and the label (i.e., the target attribute), and to correctly predict the value of the label for unseen entries.}

\nw{The first task regards the time of the day, and the label is assigned as follows: if the resources are observed between midnight and 6AM, then the label is `Night', between 6AM and 12AM `Morning', between 12AM and 6PM `Afternoon', between 6PM and midnight `Evening'.
This task answer to the following question: \textit{is there a relationship between performance and time of the day?}
The second task predicts the day of the week, with the values of the label going from `Monday' to `Sunday'.
It answers to the question: \textit{is there a relationship between performance and day of the week?}
Finally, the third task is intended to refine the previous answer and assess whether, notwithstanding the absence of significance variations across the days of the week, there is a relationship between VM performances and weekend days.
This final task is a binary classification, while the previous two are multi-class.
We employed a stratified 10-fold cross validation to train a logistic regression classification model. To confirm the obtained results, we also built five additional models using different classes of predictive algorithms: decision tree, bagging, random forest, extremely randomized trees, ada boost.}

\nw{To assess the quality of the predictions, we considered the percentage of correctly classified (accuracy) entries from the test set.
Table \ref{tab:overall_reg}  shows the mean and standard deviation of the classification accuracy aggregated by provider.
Note that a random predictor for \textit{TimeDay} would give a result of 25\%, while for \textit{DayWeek} about 14\%.
As for \textit{Weekend} prediction task, being a binary classification, the minimum baseline for accuracy is equal to 50\%.
The table also shows the results of the three classification tasks, with each of the six predictive algorithms, aggregated by VM provider.
All the algorithms scored a similar accuracy, and confirm the results obtained with logistic regression, despite the slightly lower results from decision tree and ada boost.}

\nw{Concerning both \textit{TimeDay} and \textit{Weekend}, all the algorithms performed significantly better than a random model, demonstrating the ability of ML models to extract knowledge and identify patterns where simple analyses fail.
For instance, models more likely to make correct predictions with GCP and AWS with \textit{TimeDay}, in line with the results obtained with time-series forecasts.
Instead, GCP appears much less predictable with \textit{Weekend}, which may indicate two things: there is no significant difference between the performance offered on weekends and weekdays, or there is too much variability in the offered performance across days. Given the results from time-series forecasting (i.e., that GCP is the more predictable model), we are inclined towards the first option.
At the same time, no algorithm is able to achieve very high accuracy, leaving room for discussion: VM performance appears to be predictable, but at the same time these predictions are frequently inaccurate, demonstrating that variability is still an issue with VMs, and strengthening the need for a performance prediction index, like our \indicator.}

\nw{On the other hand, accuracy for \textit{DayWeek} is very low and close to a random model, indicating the inability to assign a certain set of performance to specific days of the week.}
%However, it is worth considering that the training set is relatively small in this respect: we collected observations at intervals of 1 hour, thus 168 are needed to describe an entire week, however our data cover only around 4 weeks of observations, i.e., an insufficient amount of time for a model to learn any significant pattern.

\begin{table*}[t]
\renewcommand\arraystretch{1.4}
\setlength{\tabcolsep}{7pt}
\fontsize{8}{9.6}\selectfont
\centering
\begin{tabular}{r|c|cccccc}
\textbf{Task} & \textbf{Provider} & \textbf{Logistic Regression} & \textbf{Decision Tree} & \textbf{Bagging} & \textbf{Random Forest} & \textbf{Random Trees} & \textbf{Ada Boost} \\ \hline
\multirow{4}{*}{\textit{TimeDay}} & AWS         & 0.55$\pm$0.04 & 0.48$\pm$0.04 & \textbf{0.56$\pm$0.03} & 0.55$\pm$0.04 & 0.55$\pm$0.05 & 0.47$\pm$0.05  \\
& Azure         & \textbf{0.44$\pm$0.05} & 0.37$\pm$0.05 & 0.43$\pm$0.04 & 0.42$\pm$0.05 & 0.41$\pm$0.04 & 0.36$\pm$0.06  \\
& GCP         & 0.51$\pm$0.05 & 0.44$\pm$0.04 & 0.51$\pm$0.05 & \textbf{0.52$\pm$0.04} & 0.51$\pm$0.04 & 0.43$\pm$0.05  \\ 
& EGI         & 0.30$\pm$0.05 & 0.28$\pm$0.05 & \textbf{0.33$\pm$0.04} & 0.29$\pm$0.05 & 0.28$\pm$0.05 & 0.27$\pm$0.05  \\ \hline
\multirow{4}{*}{\textit{WeekDay}} & AWS         & 0.20$\pm$0.03 & 0.21$\pm$0.04 & 0.24$\pm$0.05 & 0.24$\pm$0.04 & \textbf{0.25$\pm$0.04} & 0.19$\pm$0.03 \\ 
& Azure         & \textbf{0.20$\pm$0.04} & 0.18$\pm$0.03 & \textbf{0.20$\pm$0.04} & 0.20$\pm$0.05 & \textbf{0.20$\pm$0.04} & 0.19$\pm$0.03 \\
& GCP         & 0.17$\pm$0.04 & 0.16$\pm$0.04 & \textbf{0.18$\pm$0.04} & 0.17$\pm$0.04 & 0.16$\pm$0.04 & 0.17$\pm$0.05 \\
& EGI         & 0.18$\pm$0.04 & 0.18$\pm$0.04 & \textbf{0.20$\pm$0.04} & 0.19$\pm$0.03 & 0.19$\pm$0.05 & 0.18$\pm$0.04 \\ \hline
\multirow{4}{*}{\textit{Weekend}} & AWS         & 0.71$\pm$0.02 & 0.65$\pm$0.05 & \textbf{0.74$\pm$0.03} & \textbf{0.74$\pm$0.03} & \textbf{0.74$\pm$0.03} & 0.68$\pm$0.03 \\
& Azure         & \textbf{0.73$\pm$0.02} & 0.66$\pm$0.05 & \textbf{0.73$\pm$0.02} & 0.72$\pm$0.03 & 0.72$\pm$0.03 & 0.68$\pm$0.05 \\
& GCP         & 0.69$\pm$0.01 & 0.59$\pm$0.05 & \textbf{0.70$\pm$0.01} & 0.68$\pm$0.03 & 0.68$\pm$0.03 & 0.62$\pm$0.05 \\
& EGI         & 0.73$\pm$0.02 & 0.65$\pm$0.05 & \textbf{0.75$\pm$0.02} & \textbf{0.75$\pm$0.02} & \textbf{0.75$\pm$0.02} & 0.68$\pm$0.04 \\ \bottomrule
\end{tabular}
\caption{\nw{Overall accuracy results for classification experiments regarding time-of-the-day, day-of-the-week and weekend predictive tasks, aggregated by VM provider. Best result for each task and each provider in bold.}}
\label{tab:overall_reg}
\end{table*}

\section{Discussion}
\label{sec-discussion}
\nw{The wide range of experiments that we run allowed us to investigate VM performance variability from different points of view and to answer our initial research questions.}

\nw{To answer \rqvi, we created \indicator, an indicator of variability for VMs.
The results obtained from analyzing 16 VMs and 4 VM providers are succinct and clear, and confirm the trend observed in the rest of the analyses presented in the paper:
AWS and GCP perform better than the other providers, especially with smaller machines (C1/C2), while Azure exhibits without any doubt the more unstable performance.
For instance, Azure A2 and A4 showed the worst performance: among 40 metrics for classes C1 and C2, Azure machines obtained unstable results on 21.
These results are also aligned with those obtained with ML predictive models, where Azure appeared the most unpredictable among the providers considered, thus highlighting a connection between performance instability and unpredictability.
Moreover, this confirms that \indicator is able to identify important performance trends that only ML analyses were capable to display.}

\nw{In summary, \indicator allows to estimate variability for a given VM, considering a large set of metrics.
Moreover, its flexibility allows to easily adjust the time frame of the considered data, permitting more refined analyses.
The results presented in this paper demonstrate that \indicator is able to clearly capture the most significant trends in VM performances.}

\begin{answerbox}{\rqvitext}
Performance variability can be analyzed by considering multiple dimensions: breadth, dispersion, speed. We introduce a \fullindicator (\indicator) that is able to quantify the variability in a given user defined time window, and identify stable and unstable VMs and providers.
\end{answerbox}

\nw{To answer \rqres, we first studied the relative standard deviation of the measured performance metrics, showing that different resources are affected by different levels of variability, even for the same provider.
Moreover, we provided an extensive study on the prediction of individual VM resources using time-series forecasting models, analyzing both the perspective of providers and classes of VMs offered.
The intuition is that the easier is to predict its resource performance, the more stable is the VM.}

\nw{Prediction tasks gave us important insights.
We show that making accurate predictions is not always possible, but more advanced models are also more accurate.
This tells us that predictable patterns are present in the collected data, although they can be often difficult to extrapolate.
Moreover, the results confirmed that not all providers behave in the same way: for instance, Azure appears much less predictable, thus more unstable.
On the other hand, there is not much difference when comparing different classes of VMs, although smaller machines appear slightly more predictable.
From the perspective of individual resources, the variability is most noticeable with Disk and Network metrics, while Memory is more predictable.}

\begin{answerbox}{\rqzero}
The results obtained by single providers on different resources show different levels of variability. For example, EGI tends to be stable especially on CPU, but very unstable on Disk.
Overall, if one accepted some variability, AWS and GCP behave in a more appropriate way, while EGI and mostly Azure do not. 
\end{answerbox}

\nw{To answer \rqiso, we first observed the correlation matrix between performance metrics of each provider, then performed a gradient analysis on them. It is important to note that our experiments aimed to test different resources independently from the another and then compare their fluctuations in a same time window. Overall, results suggest that providers are capable of managing resources in isolation.}

\nw{Our test indicates that metrics are usually not correlated among them, in some cases not even between metrics that measure the same resource, but with different benchmarks.
A higher correlation was evidenced in GCP and Azure, while most of the metrics in AWS and EGI did not appear to be correlated. This means that AWS and EGI are the two cloud providers that provide better isolation among resources. Overall, we found that only CPU and memory showed a consistent but rather weak correlation across all providers.
Moreover, the gradient analysis confirmed that there is no strong dependency among single resources, and only discovered a weak correlation  with Azure.}

\begin{answerbox}{\rqthree}
We did not find any noticeable propagation among resource degradation, except for a weak correlation among resources (particularly with GCP and Azure VMs). Thus, only a multi-metrics analysis allows one to grasp how variability affects the VMs over time.
\end{answerbox}

\nw{To answer \rqtime, we first performed a data analysis, but it was not possible to recognize any recurring temporal pattern in the performance data.
Therefore, we indirectly studied whether time can influence the performance of VM providers through three ad-hoc classification tasks.
We noticed that performance is extremely difficult to associate with a specific day of the week (\textit{WeekDay} task), regardless of the predictive model employed and the provider considered, with results on par with completely random predictions.
On the other hand, it is possible to distinguish resource performances during weekend and working days, especially with Azure and EGI (\textit{Weekend} task).
In this experiment, GCP again appears to be the more stable provider, as it shows smaller differences in performance between weekends and weekdays, therefore making classification more difficult for the predictive model.
Concerning the relationship with time of the day (\textit{TimeDay} task), AWS and GCP are clearly the more predictable models, thus resulting in more recognizable time patterns compared to Azure and EGI.
Finally, covering a variety of predictive models allows to conclude that both logistic regression and ensemble models based on bagging, random forest and randomized trees perform equally well, while decision tree and ada boost are not adequate for predicting trends with VM performance data.}

\begin{answerbox}{\rqone}
According to our data analysis, variations occur at random times; regression and ensemble models can instead identify patterns where simple analyses fail, highlighting the relation between offered performance and both time of the day and weekends.
However, we could not conclude that variations are dependent on the specific day of the week.
%There is no noticeable relationship between hour/day/week and the magnitude, duration, and speed of the variation.
\end{answerbox}

\nw{To answer \rqcost, we compared across VM providers the mean and standard deviation of the various performance metrics observed.
We noticed that in many cases (e.g., CPU EVENTS) the performance is not proportionate to the cost of the VM, suggesting that different performances do not necessarily depend on the size of the machine.
Moreover, variations in the performance are not always similar in nature across providers.}

\nw{Additionally, we computed the cost/performance ratio for different performance metrics.
in general, our experiments suggested that VM performances mainly depend on the metric itself and not on the type and size of the VM.
CPR also assessed important differences between VM providers, with AWS and Azure seeming to offer the best cost to performance ratio for smaller and bigger VMs, respectively. We discovered that small machines often offer the best cost/performance ratio. In four of the five studied metrics, the best results where obtained by machine in class C1 of GCP (1/5) and AWS (3/5). 
In general we did not observed a strong correlation between stability of performance and cost. For example, Azure VMs appear to often have the highest variability among the studied providers, but some times they are also the most convenient ones.}

\begin{answerbox}{\rqfour}
Measured values say that more expensive VMs do not always offer better performance
On average, AWS seems to be better with the smaller VMs while Azure with the bigger
machines provide the most cost-effective choices. Oftentimes, practitioners must face the trade-off between convenience and performance stability.
\end{answerbox}

\subsection{Threats to Validity}
This section summarizes the most important threats to the validity of presented results by following the structure proposed in~\cite{wohlin_empirical_research}.

\textbf{Internal Threats.} Cloud providers may update their infrastructures, change the hardware that powers their services, and implement new software features. Most of the time, all these activities are transparent to the user, but they may affect benchmark executions, and thus the obtained results.
\nw{The use of a month as time window helped us mitigate this problem. It allowed us to analyze the VMs over a significant amount of time and thus take into account possible changes and updates.
However, for validating the impossibility of predicting the day of the week, a longer observation window may be required.}

In addition, executed benchmarks may be subject to flaws or bugs. To soften this aspect, we executed well-known benchmarks (i.e., SysBench and Nench) that are widely adopted and also used in related works \cite{patterns_leitner}.

We compared VMs with different underlying CPU architectures (e.g., ARM vs Intel). However, we compared VM types with similar resources (cores and memory), cost, and purpose. In general, we observed that the underlying hardware architecture does not significantly impact the variability of VMs, whereas, as expected, their behavior depends on the interplay of several ``hidden'' factors.

\textbf{External Threats.} The number of tested providers is limited, but we selected the most popular commercial and research-oriented (in Europe) solutions to provide a wide and clear picture of the current mainstream offers. We also executed the benchmarks on a limited number of VMs, and we only picked a subset of offered VM types and sizes.
Furthermore, all the benchmarks were executed on machines located in Europe, but other (smaller) assessments \cite{ec2_performance_analysis_dejun}\cite{schad_runtime_measurements} notice that the performance also fluctuates in other regions and that depends on the regions themselves. We are convinced that selected VMs represent a common choice for many users, and also allowed us to keep the cost of these experiments under control.
More providers, regions, VM types, and instances would have helped us better understand how performance varies, but we tend to say that obtained results could qualitatively be very similar to ours.

The work only focused on VMs as computing resource since it is probably the most popular IaaS service nowadays. Needless to say, other resources (e.g., containers) and PaaS services should be tested separately to get an even better landscape of the performance of cloud infrastructures. Given the wide adoption of VMs, and given the fact that other computing resources are based on them, we are confident this can be a good starting point.

\textbf{Conclusion Threats.} We executed $10$ different benchmarks to measure $28$ different metrics in total. We also repeated the executions multiple times, every hour over a one-month period, to collect a huge amount of diverse measurements. The number of repetitions was chosen to balance total execution time and the statistical validity of obtained results. The type and number of benchmarks and the number of repetitions can be increased to widen the test area and to obtain more robust results. As drawback, execution time and cost would increase.

\section{Related work}
\label{sec-related}

Many researchers have already addressed the problem of evaluating the performance of VMs.
Some works want to test specific resources while others are interested in the performance of software applications that exploit multiple resources concurrently. These works differ for considered providers, size of employed VMs, run benchmarks, systems used to set up and collect measured values, and data aggregation and analysis techniques. The conclusions have been often partial and witness unstable cloud resources. For example, years ago the work by Iosup et al.~\cite{iosup_performance_analysis_scientific_computing_2011} highlighted that the cloud was not suitable for running scientific experiments. Similarly, Salah et al.~\cite{salah_performance}  analyzed the performance of VPS (virtual private servers) on AWS and other two cloud providers (ElasticHosts and BlueLock) and shows that the performance and its variability depend on executed benchmarks and selected provider (even if they did not monitor the performance over a period of time).

Our experiments share some key elements with the following works. Leitner and Cito~\cite{patterns_leitner} formulate 15 hypotheses on performance variations in IaaS systems. They analyze the factors that influence variation and how VM sizes can be compared. They run $3$ benchmarks to target CPU, disk, and memory speeds, and $2$ application benchmarks to measure the queries per minute on a MySQL database and the Git checkout and Java compilation times of an open source project. They collected data from four cloud providers (AWS, GCP, Azure, and IBM) to validate the hypotheses. They run both \textit{isolated} and \textit{continuous} tests. The former tests acquired and initialized a VM and executed the benchmark three times in a row, 6 times per day, over a period of one month. The latter tests acquired and initialized a VM and the benchmark was executed once every hour over 3 days. They state that cloud performance is a ``moving target'' and that the scientific community is required to periodically re-validate its understanding of the subject. Our work considered tests that are similar to their continuous ones, but for a longer period. Moreover, we executed a greater number of benchmarks, and few are similar to theirs. They used Workbench to setup the data collection while we design and develop a custom and open system. 

Gillam et al.~\cite{gillam2013fair} use different benchmarks to evaluate diverse resources: memory, CPU, disk, and network provided by four different cloud providers (AWS, Rackspace, IBM, and a private cloud installation of OpenStack). They executed a few benchmarks and say that the variability does not only depend on the provider, but it also comes from different instances of the same resource types. We executed a different set of benchmarks, on different cloud providers, using a different approach (RMT) for launching the benchmarks as described in Section \ref{sec-methodology}.

\nw{Sharma et al. \cite{sharma2016containers} explore how multi-tenancy affect the performance of VMs and containers. They analyze whether the utilization of similar or different resources by different tenants sharing a common underlying resource (e.g., a physical machine) affects higher-level performance metrics (e.g., response time). We also study whether cloud providers manage resources in isolation and we found out that the performance of different resources are weakly correlated. However, we assumed that the cloud is a black-box and we do not have access to information regarding the activities running on the same psychical resources.} 

Li et al.~\cite{li2011comparing, cloudcmp_li} present CloudCmp, a systematic comparator of the performance and cost of cloud providers. They consider offered services, devise a set of metrics related to application performance, and propose a set of tools for measuring them. The metrics measure instance, storage, and network efficiency, performance/cost ratio, scaling speed, storage consistency, and WAN latency. They executed benchmarks for computing resources, persistent storage, and network. Their approach aims to devise a tool that exploits collected metrics to help users select the most suitable service. Our work in contrast is more focused on collecting metrics to investigate whether one can predict VM performance and study its variability across different providers. Also Samreen et al. \cite{samreen_daleel} propose a framework, called \textit{Daleel}, to support adaptive decision making in cloud environments. While their work focuses on the use of machine learning techniques, they also analyzed the variability of AWS EC2 VMs (by means of a simple application and only for a week). The results say that the performance is not always proportional to the cost of the VM, and that the variability depends on the VM type and on the time of the execution.

%He et al.~\cite{statistics_he} propose a performance testing methodology called PT4Cloud, a statistical approach that employs the likelihood theory and the bootstrap method to identify proper stop conditions for creating performance distributions. The approach is evaluated on AWS and Chameleon, another research-oriented cloud provider, with diverse benchmarks. PT4Cloud was able to provide an average 95\% accuracy with a 62\% reduction of test runs. Our work starts a step behind, and analyses data to help us understand whether we need such statistical approaches to create performance distributions, and in case, we want to study the hypotheses and conditions that must hold to apply them successfully. 

Other works propose models to solve the cloud performance variability problem. For example, Ardagna et al.~\cite{ardagna2014quality} survey current approaches for workload and system modeling for assessing the quality of service (QoS) of cloud applications to guarantee performance, availability, and reliability. The goal is to provide suitable modeling means without any reference to concrete measured values.
Casale and Tribastone~\cite{modelling_casale} propose the \textit{blending} solution for the analysis of queueing networks-based performance models that include exogenous variability, modelled as a continuous-time Markov chain. Again, this work proposes a modelling technique that can save computational time when compared to simulation. Compared to these approaches, we proposed to measure cloud variability using a novel indicator (VI) and to predict future performance using ML methods. 

Podolskiy et al.~\cite{podolskiy_iaas_autoscaling} evaluate the auto-scaling capabilities of AWS, Azure, and GCP VMs as means to (also) cope with their variability.
Gesvindr and Buhnova~\cite{performance_challenges_gesvindr} analyze the performance, scalability, elasticity, and availability of cloud applications and propose best practices for optimizing them. Among the given advices, they suggest to collect throughput metrics to identify the suitability of cloud providers and to select low response-time services among the offered ones. Our work moves a step forward and \indicator is a simple indicator created to characterize VMs and may be used in addition to proposed practices.

Some other works focus on the repeatability of cloud-based experiments. Abedi et al.~\cite{repeatable_experiments_abedi} discuss the most commonly used approaches for comparing performance measurements in cloud environments and show that there exist flaws in methodologies that may lead to erroneous conclusions. They present a methodology for executing repeatable experiments in environments where conditions change and are not under user control. 
More recently, He et al. \cite{DBLP:conf/kbse/HeLLLK021} notice how the high variability of cloud platforms can affect the statistical relevance of experiments run in such environments. To solve this problem they propose Metior, a solution that allows to reduce the errors on performance testing performed in the cloud. Their tool repeats the execution of the application under test intermittently and on two consecutive days. If the data obtained on the two days are comparable the results are accurate, otherwise the procedure is repeated in the following days until matching data are found. The comparison is performed using \textit{block bootstrapping}, a random sampling technique for time-series that removes biases and outliers.
We consider these approaches complementary to our work: on the one hand our analysis confirms that the variability of cloud platforms may affect significantly the execution of running applications on diverse metrics and dimensions, on the other our \fullindicator can be exploited by novel tools to understand if and to what extent cloud performance is stable.

\nw{Some few works analyzed cloud performance using ML techniques. For example, Grohmann et al. \cite{grohmann2019monitorless} exploits ML to predict the response time of running applications from low-level infrastructural metrics. Similarly, Rahman et al. \cite{rahman2019predicting} studies the changes of the monitored response time given induced interferences such as co-locating multiple VMs on a single physical machine (i.e., multi-tenant resource sharing). Compared to this work, we exploited ML techniques to study performance variability of VM considering all the resources and different cloud providers. In particular, we searched for patterns in time to forecast the future performance of the cloud platform and not how low-level metrics impact high-level ones (such as response time).}

Papadopoulos et al.~\cite{papadopoulos_principles_reproducible_evaluation} investigate how to measure and report performance in the cloud and how well the cloud research community is already doing it. They propose a set of eight methodological principles that combine best-practices from nearby fields with concepts only applicable to the cloud. They also report on a systematic literature review to analyze whether the practice of reporting cloud performance measurements follows the proposed principles. We executed the experiments following the general principles proposed by the authors. We repeated the experiments multiple times (P1), with different workloads and by randomizing the execution (P2), we describe hardware and software setup properly (P3), used data are open and available (P4), we also provide aggregated data about measured performance (P5) and a statistical evaluation (P6), we specify the measured units for every benchmark in graphs and tables (P7) and the used cost model. We also reported the cost model adopted for the experiments, used resources, and the costs of benchmarked providers (P8).
All these suggestions and principles helped us conceive a proper methodology and increase the soundness of our results.

\section{Conclusions and Future Work}
\label{sec-conclusions}
This paper presents a multi-faceted analysis of the (performance) variability of four common VM classes offered by well-known cloud providers (AWS, GCP, Azure, and EGI). The analysis considers $28$ different measures, and this is the widest assessment we are aware of. Our analyses provide some key insights. For example, we discovered that resources have different variability within the same provider and multiple metrics are required to fully understand to what extent the environment is stable. However, in general results are complex and difficult to fully understand given that cloud providers abstract away multiple layers of management. For this reason, we introduce a relatively simple variability indicator, called \indicator, that is able to summarize and quantify the variability of cloud infrastructures.
\nw{Moreover, we also analyzed the performance of VM using ML techniques that were able to discover latent time patterns.}
Overall, our experiments highlight that AWS provides the most stable infrastructure, GCP obtained comparable but slightly worse results, while Azure and EGI appear to be more noisy and unstable.
\nw{Our future plans involve expanding the observation window to encompass multiple months. This extension will enable us to leverage a broader dataset and develop more robust predictive models with enhanced capabilities. By incorporating an extended timeframe into our analysis, we aim to strengthen the accuracy and effectiveness of the computed predictions.}

\bibliography{bib}

\begin{thebibliography}{10}
\providecommand \doibase [0]{http://dx.doi.org/}%

\bibitem{armbrust2010view}
Armbrust M, Fox A, Griffith R, et al. {A View of Cloud Computing}. {\it
  Communications of the ACM} 2010\string; 53(4)\string: 50--58.

\bibitem{caas_what_is}
What is {Containers} as a service ({CaaS})?.
  https://www.ibm.com/services/cloud/containers-as-a-service;  2022.

\bibitem{serverless_what_is}
What is {Serverless Computing}?. https://www.ibm.com/cloud/learn/serverless;
  2022.

\bibitem{DBLP:journals/jss/TianTL20}
Tian Y, Tian J, Li N. Cloud reliability and efficiency improvement via failure
  risk based proactive actions. {\it J. Syst. Softw.} 2020\string; 163\string:
  110524.

\bibitem{performance_challenges_gesvindr}
Gesvindr D, Buhnova B. Performance Challenges, Current Bad Practices, and Hints
  in PaaS Cloud Application Design. {\it SIGMETRICS Perform. Eval. Rev.}
  2016\string; 43(4)\string: 3–12.
\newblock \href {\doibase 10.1145/2897356.2897358} {doi:
  10.1145/2897356.2897358}

\bibitem{qos_cloud_computing_Ardagna}
Ardagna D, Casale G, Ciavotta M, P{\'e}rez JF, Wang W. Quality-of-service in
  cloud computing: modeling techniques and their applications. {\it Journal of
  Internet Services and Applications} 2014\string; 5(1)\string: 11.
\newblock \href {\doibase 10.1186/s13174-014-0011-3} {doi:
  10.1186/s13174-014-0011-3}

\bibitem{patterns_leitner}
Leitner P, Cito J. Patterns in the Chaos—A Study of Performance Variation and
  Predictability in Public IaaS Clouds. {\it ACM Trans. Internet Technol.}
  2016\string; 16(3).
\newblock \href {\doibase 10.1145/2885497} {doi: 10.1145/2885497}

\bibitem{azure_status}
Azure status. https://status.azure.com/en-us/status;  2022.

\bibitem{aws_status}
{AWS} {Service} {Health} {Dashboard}. https://status.aws.amazon.com/;  2022.

\bibitem{azure_disks}
Select a disk type for {Azure} {IaaS} {Linux} {VMs} - managed disks - {Azure}
  {Linux} {Virtual} {Machines}.
  https://docs.microsoft.com/en-us/azure/virtual-machines/linux/disks-types;
  2022.

\bibitem{aws_disks}
Amazon {EBS} volume types - {Amazon} {Elastic} {Compute} {Cloud}.
  https://docs.aws.amazon.com/AWSEC2/latest/UserGuide/ebs-volume-types.html;
  2022.

\bibitem{gcp_disks}
Storage options {\textbar} {Compute} {Engine} {Documentation}.
  https://cloud.google.com/compute/docs/disks;  2022.

\bibitem{cloudcmp_li}
Li A, Yang X, Kandula S, Zhang M. CloudCmp: Comparing Public Cloud Providers.
  In:   {\it Proceedings of the 10th ACM SIGCOMM Conference on Internet
  Measurement}IMC ’10. Association for Computing Machinery. ; 2010; New York,
  NY, USA\string: 1–14

\bibitem{cloud_benchmark_suite_scheuner}
Scheuner J, Leitner P. A Cloud Benchmark Suite Combining Micro and Applications
  Benchmarks. In:   {\it Companion of the 2018 ACM/SPEC International
  Conference on Performance Engineering}ICPE ’18. Association for Computing
  Machinery. ; 2018; New York, NY, USA\string: 161–166

\bibitem{microbenchmarking_laaber}
Laaber C, Scheuner J, Leitner P. Software Microbenchmarking in the Cloud. How
  Bad is it Really?. {\it Empirical Software Engineering} 2019\string; 24.
\newblock \href {\doibase 10.1007/s10664-019-09681-1} {doi:
  10.1007/s10664-019-09681-1}

\bibitem{ec2_performance_analysis_dejun}
Dejun J, Pierre G, Chi CH. EC2 Performance Analysis for Resource Provisioning
  of Service-Oriented Applications. In:   {\it Service-Oriented Computing.
  ICSOC/ServiceWave 2009 Workshops}Springer Berlin Heidelberg. ; 2010; Berlin,
  Heidelberg\string: 197--207.

\bibitem{repeatable_experiments_abedi}
Abedi A, Brecht T. Conducting Repeatable Experiments in Highly Variable Cloud
  Computing Environments. In:   {\it Proceedings of the 8th ACM/SPEC on
  International Conference on Performance Engineering}ICPE ’17. Association
  for Computing Machinery. ; 2017; New York, NY, USA\string: 287–292

\bibitem{statistics_he}
He S, Manns G, Saunders J, Wang W, Pollock L, Soffa ML. A Statistics-Based
  Performance Testing Methodology for Cloud Applications. In:   {\it
  Proceedings of the 2019 27th ACM Joint Meeting on European Software
  Engineering Conference and Symposium on the Foundations of Software
  Engineering}ESEC/FSE 2019. ACM. ; 2019\string: 188–199.

\bibitem{papadopoulos_principles_reproducible_evaluation}
Papadopoulos A, Versluis L, Bauer A, et al. Methodological Principles for
  Reproducible Performance Evaluation in Cloud Computing. {\it IEEE
  Transactions on Software Engineering} 5555(01)\string: 1-1.
\newblock \href {\doibase 10.1109/TSE.2019.2927908} {doi:
  10.1109/TSE.2019.2927908}

\bibitem{dickey1979distribution}
Dickey DA, Fuller WA. Distribution of the estimators for autoregressive time
  series with a unit root. {\it Journal of the American statistical
  association} 1979\string; 74(366a)\string: 427--431.

\bibitem{lutkepohl2005new}
L{\"u}tkepohl H. {\it New introduction to multiple time series analysis}.
\newblock Springer .
\newblock 2005

\bibitem{box2015time}
Box GE, Jenkins GM, Reinsel GC, Ljung GM. {\it Time series analysis:
  forecasting and control}.
\newblock John Wiley \& Sons .
\newblock 2015.

\bibitem{durbin2012time}
Durbin J, Koopman SJ. {\it Time series analysis by state space methods}. 38.
\newblock OUP Oxford .
\newblock 2012.

\bibitem{hyndman2006another}
Hyndman RJ, Koehler AB. Another look at measures of forecast accuracy. {\it
  International journal of forecasting} 2006\string; 22(4)\string: 679--688.

\bibitem{wohlin_empirical_research}
Wohlin C, H{\"o}st M, Henningsson K. {\it Empirical Research Methods in Web and
  Software Engineering}\string: 409--430; Springer Berlin Heidelberg .
\newblock 2006

\bibitem{schad_runtime_measurements}
Schad J, Dittrich J, Quian\'{e}-Ruiz JA. Runtime Measurements in the Cloud:
  Observing, Analyzing, and Reducing Variance. {\it Proc. VLDB Endow.}
  2010\string; 3(1–2)\string: 460–471.
\newblock \href {\doibase 10.14778/1920841.1920902} {doi:
  10.14778/1920841.1920902}

\bibitem{iosup_performance_analysis_scientific_computing_2011}
{Iosup} A, {Ostermann} S, {Yigitbasi} MN, {Prodan} R, {Fahringer} T, {Epema} D.
  Performance Analysis of Cloud Computing Services for Many-Tasks Scientific
  Computing. {\it IEEE Transactions on Parallel and Distributed Systems}
  2011\string; 22(6)\string: 931-945.
\newblock \href {\doibase 10.1109/TPDS.2011.66} {doi: 10.1109/TPDS.2011.66}

\bibitem{salah_performance}
Salah K, Al-Saba M, Akhdhor M, Shaaban O, Buhari M. Performance evaluation of
  popular Cloud IaaS providers. In:   {\it 2011 International Conference for
  Internet Technology and Secured Transactions}IEEE. ; 2011\string: 345--349.

\bibitem{gillam2013fair}
Gillam L, Li B, O’Loughlin J, Tomar APS. Fair benchmarking for cloud
  computing systems. {\it Journal of Cloud Computing: Advances, Systems and
  Applications} 2013\string; 2(1)\string: 6.

\bibitem{sharma2016containers}
Sharma P, Chaufournier L, Shenoy P, Tay Y. Containers and virtual machines at
  scale: A comparative study. In:   {\it Proceedings of the 17th international
  middleware conference}ACM. ; 2016\string: 1--13.

\bibitem{li2011comparing}
Li A, Yang X, Kandula S, Zhang M. Comparing public-cloud providers. {\it IEEE
  Internet Computing} 2011\string; 15(2)\string: 50--53.

\bibitem{samreen_daleel}
Samreen F, Elkhatib Y, Rowe M, Blair GS. Daleel: Simplifying cloud instance
  selection using machine learning. In:   {\it NOMS 2016 - 2016 IEEE/IFIP
  Network Operations and Management Symposium}IEEE. ; 2016\string: 557-563

\bibitem{ardagna2014quality}
Ardagna D, Casale G, Ciavotta M, P{\'e}rez JF, Wang W. Quality-of-service in
  cloud computing: modeling techniques and their applications. {\it Journal of
  Internet Services and Applications} 2014\string; 5(1)\string: 11.

\bibitem{modelling_casale}
Casale G, Tribastone M. Modelling Exogenous Variability in Cloud Deployments.
  {\it SIGMETRICS Perform. Eval. Rev.} 2013\string; 40(4)\string: 73–82.
\newblock \href {\doibase 10.1145/2479942.2479951} {doi:
  10.1145/2479942.2479951}

\bibitem{podolskiy_iaas_autoscaling}
Podolskiy V, Jindal A, Gerndt M. IaaS Reactive Autoscaling Performance
  Challenges. In:   {\it 2018 IEEE 11th International Conference on Cloud
  Computing (CLOUD)}IEEE. ; 2018\string: 954-957

\bibitem{DBLP:conf/kbse/HeLLLK021}
He S, Liu T, Lama P, Lee J, Kim IK, Wang W. Performance Testing for Cloud
  Computing with Dependent Data Bootstrapping. In:   {\it 36th {IEEE/ACM}
  International Conference on Automated Software Engineering, {ASE}}{IEEE}. ;
  2021\string: 666--678.

\bibitem{grohmann2019monitorless}
Grohmann J, Nicholson PK, Iglesias JO, Kounev S, Lugones D. Monitorless:
  Predicting performance degradation in cloud applications with machine
  learning. In:   {\it Proceedings of the 20th international middleware
  conference}ACM. ; 2019\string: 149--162.

\bibitem{rahman2019predicting}
Rahman J, Lama P. Predicting the end-to-end tail latency of containerized
  microservices in the cloud. In:   {\it 2019 IEEE International Conference on
  Cloud Engineering (IC2E)}IEEE. ; 2019\string: 200--210.

\end{thebibliography}

\clearpage

% \appendix
% \label{appendix}
% \input{sections/appendix}

% \begin{IEEEbiography}[{\includegraphics[width=1in,height=1.25in,clip,keepaspectratio]{baresi}}]{Luciano Baresi} is a full professor at the Politecnico di Milano.  Luciano was visiting professor at University of Oregon (USA) and visiting researcher at University of Paderborn (Germany). His research interests are in the broad area of software engineering and include formal approaches for modeling and specification languages, distributed systems, service-based applications and mobile, self-adaptive, and pervasive software systems. 
% \end{IEEEbiography}
% \vskip 0pt plus -1fil
% \begin{IEEEbiography}[{\includegraphics[width=1in,height=1.25in,clip,keepaspectratio]{quattrocchi}}]{Giovanni Quattrocchi}
% received his Ph.D. in Computer Engineering in 2018 from Politecnico di Milano, where he is currently a post-doc researcher. He was a visiting researcher at University of California San Diego and Imperial College London.  His research interests include self-adaptive systems, software architectures, distributed systems, performance analysis, and mobile and edge computing.
% \end{IEEEbiography}
% \vskip 0pt plus -1fil
% \begin{IEEEbiography}[{\includegraphics[width=1in,height=1.25in,clip,keepaspectratio]{rasi}}]{Nicholas Rasi}
% received his master degree in Computer Science and Engineering in 2019 from Politecnico di Milano where is currently a research assistant. His research interests include heterogeneous systems, self-adaptive systems, service-oriented architectures and performance analysis.
% \end{IEEEbiography}

\end{document}